\documentclass[a4paper,twocolumn,english,showpacs,superscriptaddress,prc]{revtex4-1}
\usepackage[T1]{fontenc}
\usepackage[latin9]{inputenc}
\usepackage{color}
\usepackage{babel}
\usepackage{float}
\usepackage{amsmath}
\usepackage{amssymb}
\usepackage{graphicx}
\usepackage[unicode=true,pdfusetitle,
 bookmarks=true,bookmarksnumbered=false,bookmarksopen=false,
 breaklinks=false,pdfborder={0 0 1},backref=false,colorlinks=true]
 {hyperref}
\hypersetup{
 citecolor=blue,linkcolor=blue}

\newcommand\ins[1]{\textcolor{black}{#1}}

\makeatletter

\pdfpageheight\paperheight
\pdfpagewidth\paperwidth

\makeatother

\begin{document}
\title{Elliptic flow of hadrons in equal-velocity quark combination mechanism in relativistic heavy-ion collisions}
\author{Jun Song }
\address{Department of Physics, Jining University, Shandong 273155, China}
\author{Hai-hong Li}
\address{Department of Physics, Jining University, Shandong 273155, China}
\author{Feng-lan Shao}
\email{shaofl@mail.sdu.edu.cn}

\address{School of Physics and Physical Engineering, Qufu Normal University,
Shandong 273165, China}

\begin{abstract}
    We apply a quark combination model with equal-velocity combination (EVC) approximation to study the elliptic flow ($v_{2}$) of hadrons in heavy-ion collisions in a wide collision energy range ($\sqrt{s_{NN}}=$ 27 - 5020 GeV). Utilizing the simple relationship between $v_{2}$ of hadrons and those of quarks under EVC, \ins{we find that} $v_{2}$ of up/down quarks \ins{obtained} by experimental data of proton is consistent with that obtained by data of $\Lambda$ and $\Xi$. $v_{2}$ of strange quarks obtained by data of $\Omega$ is consistent with that obtained by data of $\Lambda$ and $\Xi$, and at RHIC energies it is also consistent with that obtained by data of $\phi$.  This means that $v_{2}$ of these hadrons have a common quark-level source. \ins{Using data of $D^0$,} we \ins{obtain} $v_{2}$ of charm quarks \ins{with $p_T\lesssim 6$ GeV/c}.  We find that \ins{under EVC} charm quark dominates $v_{2}$ of $D$ mesons at low $p_{T}$ but light-flavor quarks significantly contribute to $v_{2}$ of $D$ mesons in the range $3\lesssim p_{T}\lesssim8$ GeV/c. We predict $v_{2}$ of charmed baryons $\Lambda_{c}^{+}$ and $\Xi_{c}^{0}$ which show a significant enhancement at intermediate $p_{T}$ due to the double contribution of light-flavor quarks. The properties of the obtained quark $v_{2}$ \ins{under EVC} are studied and \ins{a} regularity for $v_{2}$ of quarks as the function of $p_{T}/m$ is found.
\end{abstract}
\maketitle

\section{Introduction}

In non-central heavy-ion collisions, momentum distributions of the
produced hadrons are anisotropic in the transverse plane perpendicular
to the beam direction \cite{Ollitrault:1992bk}.  The elliptic flow
($v_{2}$) is the second harmonic coefficient of Fourier expansion
for the azimuthal distribution of particle transverse momentum and
denotes the asymmetry between $x$ and $y$ components of particle
transverse momentum. $v_{2}$ of hadron carries important information
on the degree of initial thermalization, the pressure gradients, the
equation of state, and the hadronization for the created quark matter
\cite{Ollitrault:1992bk,Sorge:1998mk,Teaney:2000cw,Huovinen:2001cy,Adare:2006ti}.

In non-central heavy-ion collisions at RHIC and LHC energies, the
data of $v_{2}$ for light-flavor hadrons as the function of transverse
momentum ($p_{T}$) exhibit a number-of-constituent quark scaling
(NCQ) property \cite{Adare:2006ti,Adamczyk:2015ukd,Abelev:2014pua,Acharya:2018zuq,Adamczyk:2013gw}.
As $v_{2}$ and $p_{T}$ of identified hadrons are divided by the
number of constituent quarks (2 for meson, 3 for baryon), the scaled
data of different hadrons approximately follow a common tendency.
Replacing $p_{T}$ by the transverse kinetic energy $\textrm{KET}=\sqrt{p_{T}^{2}+m^{2}}-m$
in the horizontal axis, NCQ looks better. Such a NCQ is expected in
quark recombination/coalescence model \cite{Molnar:2003ff,Fries:2003kq,Fries:2003vb,Greco:2003xt,Kolb:2004gi,Minissale:2015zwa,Oh:2009zj,Plumari:2017ntm}.
The preliminary data of $D^{0}$ in Au+Au collisions at $\sqrt{s_{NN}}=$
200 GeV \cite{Singha:2018cdj} seemingly also follow this NCQ.

In our recent works, we found a constituent quark number scaling property
exhibited in the $p_{T}$ spectra of identified hadrons in high multiplicity
$pp$, $p$-Pb and AA collisions \cite{Song:2017gcz,Zhang:2018vyr,Song:2019sez}.
This property is the direct consequence of the equal-velocity combination
(EVC) of constituent quarks and antiquarks at hadronization. We further
demonstrated that EVC can self-consistently explain the data of $p_{T}$
spectra for different identified hadrons in high multiplicity $pp$
and $p$-Pb collisions at LHC energies \cite{Zhang:2018vyr,Gou:2017foe,Li:2017zuj,Song:2018tpv}
and in heavy-ion collisions in a wide collision energy range \cite{song2020strange}.
Taking advantage of rich data for hadronic $v_{2}$ in heavy-ion collisions,
in particular, at RHIC BES energies \cite{Adamczyk:2015ukd,Adamczyk:2015fum,Acharya:2018zuq,Zhu:2018zqo,Abelev:2014pua},
it is interesting to further study whether this EVC mechanism works
for $v_{2}$ of hadrons or not.

Because of different constituent mass such as $m_{d}\approx m_{u}\approx0.3$
GeV, $m_{s}\approx0.5$ GeV and $m_{c}=1.5$ GeV, quarks of different
flavors under EVC will contribute different momenta (proportional
to quark mass) to the formed hadron. Therefore, $p_{T}/2$ for meson
and $p_{T}/3$ for baryon in the aforementioned NCQ operation on $v_{2}$
data of light-flavor hadrons can not exactly denote the transverse
momentum of each quark in the hadron containing different quark flavors.
Instead, we should divide the momentum of hadron into different pieces
so as to obtain the actual momentum of the quark at hadronization
and the actual relationship between quark transverse momentum and
its $v_{2}$. Therefore, applying the EVC to study hadronic $v_{2}$
data will bring new results for $v_{2}$ of quarks at hadronization.

In this paper, we apply the quark combination model (QCM) with EVC
to systematically study the $v_{2}$ of identified hadrons and extract
the $v_{2}$ of quarks at hadronization using the available data of
identified hadrons in heavy-ion collisions in a wide collision energy
range ($\sqrt{s_{NN}}=$ 27GeV - 5.02TeV) \cite{Adamczyk:2015ukd,Adamczyk:2015fum,Acharya:2018zuq,Zhu:2018zqo,Abelev:2014pua}.
We distinguish at first the $v_{2}$ of strange quarks from that of
up/down quarks. We extract them separately from the data of different
hadrons and study the consistency of results from different extraction
channels as the test of EVC mechanism.\emph{ }With the obtained $v_{2}$
of light-flavor quarks, we further obtain the $v_{2}$ of charm quarks
from the data of $D^{0}$. We study the dominant ingredient for $v_{2}$
of $D$ mesons at different $p_{T}$ and predict $v_{2}$ of charmed
baryons $\Lambda_{c}^{+}$ and $\Xi_{c}^{0}$. Finally, we discuss
the properties for the extracted $v_{2}$ of up/down, strange and
charm quarks.

The paper is organized as follows. In Sec. \ref{sec:qcm}, we derive
the $v_{2}$ of hadrons in QCM with EVC. In Sec. \ref{sec:v2_RHIC}
and Sec. \ref{sec:v2_LHC}, we apply formulas of EVC to decompose
the data of hadron $v_{2}$ at RHIC and LHC energies into $v_{2}$
of up/down quarks and that of strange quarks at hadronization. We
study the consistency for results obtained from different extraction
channels. In Sec. \ref{sec:v2_charm}, we study the $v_{2}$ of charm
quarks extracted from $D^{0}$ data at LHC and RHIC energies. In Sec.
\ref{sec:quark_v2_property}, we study properties of the extracted
$v_{2}$ of quarks. The summary and outlook are given at last.

\section{hadronic $v_{2}$ in QCM with EVC \label{sec:qcm}}
In this section, we apply a quark combination model with equal-velocity
combination (EVC) approximation \cite{Song:2017gcz,Gou:2017foe}
to study the production of identified hadrons in the two-dimension
transverse plane at mid-rapidity. Here, we define the distribution
function $f_{h}\left(p_{T},\varphi\right)\equiv dN_{h}/dp_{T}d\varphi$
where $\varphi$ is the azimuthal angle. The distribution function
of hadron under EVC is simply the product of those of quarks and/or
antiquarks, i.e.,

\begin{align}
f_{M_{i}}\left(p_{T},\varphi\right) & =\kappa_{M_{i}}f_{q_{1}}\left(x_{1}p_{T},\varphi\right)f_{\bar{q}_{2}}\left(x_{2}p_{T},\varphi\right),\label{eq:fptphi_mi}\\
f_{B_{i}}\left(p_{T},\varphi\right) & =\kappa_{B_{i}}f_{q_{1}}\left(x_{1}p_{T},\varphi\right)f_{q_{2}}\left(x_{2}p_{T},\varphi\right)f_{q_{3}}\left(x_{3}p_{T},\varphi\right).\label{eq:fptphi_bi}
\end{align}
Under EVC, the quark and/or antiquark have the same direction ($\varphi$)
as the hadron and take a given fraction of $p_{T}$ of the hadron.
Because of $p_{i}=m_{i}\gamma v\propto m_{i}$ at equal velocity,
the momentum fraction $x_{1,2}=m_{1,2}/\left(m_{1}+m_{2}\right)$
for meson with $x_{1}+x_{2}=1$, and $x_{1,2,3}=m_{1,2,3}/\left(m_{1}+m_{2}+m_{3}\right)$
for baryon with $x_{1}+x_{2}+x_{3}=1$. $m_{i}$ is the constituent
mass of quark $q_{i}$. $\kappa_{M_{i}}$ and $\kappa_{B_{i}}$ are
coefficients independent of $p_{T}$ and $\varphi$ but can be dependent
on the hadron species and system size. Their expressions can be found
in \cite{Gou:2017foe} and are not shown here since $\kappa_{M_{i}}$
and $\kappa_{B_{i}}$ are irrelevant to the derivation of $v_{2}$.

The quark distribution function can be written in the following form

\begin{equation}
f_{q}\left(p_{T},\varphi\right)=f_{q}\left(p_{T}\right)\left[1+2\sum_{n=1}^{\infty}v_{n,q}\left(p_{T}\right)\cos n\varphi\right],\label{eq:fqptphi}
\end{equation}
where we denote $f_{q}\left(p_{T}\right)\equiv dN_{q}/dp_{T}$ as
the $\varphi$-independent $p_{T}$ distribution function. The $\varphi$
dependence part is expressed as usual in terms of the Fourier series
and the harmonic coefficient is defined as
\begin{equation}
v_{n,q}\left(p_{T}\right)=\frac{\int f_{q}\left(p_{T},\varphi\right)\cos n\varphi\,d\varphi}{\int f_{q}\left(p_{T},\varphi\right)d\varphi}.
\end{equation}

In this paper, we study the second harmonic coefficient $v_{2}$ of
hadrons. Using Eqs. (\ref{eq:fptphi_mi})-(\ref{eq:fqptphi}), we
obtain for meson $M_{i}(q_{1}\bar{q}_{2})$

\begin{align}
 & v_{2,M_{i}}\left(p_{T}\right)=\frac{\int d\varphi\cos2\varphi f_{M_{i}}\left(p_{T},\varphi\right)}{\int d\varphi f_{M_{i}}\left(p_{T},\varphi\right)}\nonumber \\
 & =\frac{1}{\mathcal{N}_{M_{i}}}\Bigl[v_{2,q_{1}}+v_{2,\bar{q}_{2}}\nonumber \\
 & \,\,\,\,+\sum_{n,m=1}^{\infty}v_{n,q_{1}}v_{m,\bar{q}_{2}}\left(\delta_{2,m+n}+\delta_{n,2+m}+\delta_{m,2+n}\right)\Bigr]
\end{align}
 with
\begin{equation}
\mathcal{N}_{M_{i}}=1+2\sum_{n=1}^{\infty}v_{n,q_{1}}v_{n,\bar{q}_{2}}.
\end{equation}
Here, we use the abbreviation $v_{2,q_{1}}$for $v_{2,q_{1}}\left(x_{1}p_{T}\right)$
and $v_{2,\bar{q}_{2}}$ for $v_{2,\bar{q}_{2}}\left(x_{2}p_{T}\right)$.
 For baryon $B_{i}\left(q_{1}q_{2}q_{3}\right)$, we have
 \begin{widetext}
\begin{align}
v_{2,B_{i}}\left(p_{T}\right) & =\frac{1}{\mathcal{N}_{B_{i}}}\Biggl\{ v_{2,q_{1}}+v_{2,q_{2}}+v_{2,q_{3}}+\sum_{n,m=1}^{\infty}\left(v_{n,q_{1}}v_{m,q_{2}}+v_{n,q_{1}}v_{m,q_{3}}+v_{n,q_{2}}v_{m,q_{3}}\right)\left(\delta_{m,n+2}+\delta_{n,m+2}+\delta_{m+n,2}\right)\nonumber \\
 & +\sum_{n,m,k=1}^{\infty}v_{n,q_{1}}v_{m,q_{2}}v_{k,q_{3}}\left(\delta_{k,m+n+2}+\delta_{k,m+n-2}+\delta_{n,k+m-2}+\delta_{n,k+m+2}+\delta_{m,k+n-2}+\delta_{m,k+n+2}\right)\Biggr\}
\end{align}
with
\begin{equation}
\mathcal{N}_{B_{i}}=1+2\sum_{n=1}^{\infty}\left(v_{n,q_{1}}v_{n,q_{2}}+v_{n,q_{1}}v_{n,q_{3}}+v_{n,q_{2}}v_{n,q_{3}}\right)+2\sum_{n,m,k=1}^{\infty}v_{n,q_{1}}v_{m,q_{2}}v_{k,q_{3}}\left(\delta_{k,m+n}+\delta_{n,k+m}+\delta_{m,k+n}\right),
\end{equation}
 \end{widetext}
where we use the abbreviation $v_{2,q_{j}}$for $v_{2,q_{j}}\left(x_{j}p_{T}\right)$
with $j=1,2,3$.

Since the data of hadronic $v_{1}$ at mid-rapidity \cite{Abelev:2008jga,Abelev:2013cva}
are only about $v_{1,h}\lesssim10^{-3}$, $v_{1}$ of quarks should
be very small and therefore can be safely neglected. In addition,
according to NCQ estimation of the 2-4th flow of quarks $v_{2/3/4,q}\sim10^{-2}$\cite{Adare:2006ti,Adamczyk:2015ukd,Abelev:2014pua,Acharya:2018zuq,Adamczyk:2013gw},
we can neglect the influence of high order terms $(v_{n,q})^{2,3}$
in $\mathcal{N}_{M_{i}}$ and $\mathcal{N}_{B_{i}}$, and obtain

\begin{align}
 & v_{2,M_{i}}\left(p_{T}\right)\nonumber \\
 & \approx v_{2,q_{1}}\left(1+\sum_{n=2}^{\infty}\frac{v_{n,q_{1}}}{v_{2,q_{1}}}v_{n+2,\bar{q}_{2}}\right)\nonumber \\
 & +v_{2,\bar{q}_{2}}\left(1+\sum_{n=2}^{\infty}\frac{v_{n,\bar{q}_{2}}}{v_{2,\bar{q}_{2}}}v_{n+2,q_{1}}\right),\label{eq:Mv2_mid}
\end{align}
and
\begin{align}
 & v_{2,B_{i}}\left(p_{T}\right)\nonumber \\
 & \approx v_{2,q_{1}}\left[1+\sum_{n=2}^{\infty}\frac{v_{n,q_{1}}}{v_{2,q_{1}}}\left(v_{n+2,q_{2}}+v_{n+2,q_{3}}\right)\right]\nonumber \\
 & +v_{2,q_{2}}\left[1+\sum_{n=2}^{\infty}\frac{v_{n,q_{2}}}{v_{2,q_{2}}}\left(v_{n+2,q_{1}}+v_{n+2,q_{3}}\right)\right]\nonumber \\
 & +v_{2,q_{3}}\left[1+\sum_{n=2}^{\infty}\frac{v_{n,q_{3}}}{v_{2,q_{3}}}\left(v_{n+2,q_{1}}+v_{n+2,q_{2}}\right)\right].\label{eq:Bv2_mid}
\end{align}
Here we split the $v_{2}$ of meson into two parts and that of baryon
into three parts. Each part is $v_{2}$ of constituent quark multiplying
a term containing the small correction from higher-order harmonic
flows.

The magnitude of the correction is a few percentages because of $v_{4,q}\sim10^{-2}$
as mentioned above \cite{Adare:2006ti,Adamczyk:2015ukd,Abelev:2014pua,Acharya:2018zuq,Adamczyk:2013gw}.
Higher-order harmonic flows are often unavailable at present and their
influence is usually expected to be not larger than these lower-order
harmonic flows. In addition, the difference among $v_{n,q}$ of different
quark flavors is usually much (about one order of magnitude) lower
than absolute value of $v_{n,q}$. Therefore, the relative difference
among terms in brackets in Eq. (\ref{eq:Mv2_mid}) and that among
these in Eq. (\ref{eq:Bv2_mid}) should be very small ($\lesssim10^{-2}$).
Eqs.(\ref{eq:Mv2_mid}) and (\ref{eq:Bv2_mid}) can be thus expressed
approximately as the simplest form

\begin{align}
v_{2,M_{i}}\left(p_{T}\right) & =v_{2,q_{1}}\left(x_{1}p_{T}\right)+v_{2,\bar{q}_{2}}\left(x_{2}p_{T}\right),\label{eq:Mv2_final_short}\\
v_{2,B_{i}}\left(p_{T}\right) & =v_{2,q_{1}}\left(x_{1}p_{T}\right)+v_{2,q_{2}}\left(x_{2}p_{T}\right)+v_{2,q_{3}}\left(x_{3}p_{T}\right).\label{eq:Bv2_final_short}
\end{align}

\section{quark $v_{2}$ at RHIC\label{sec:v2_RHIC}}

In this section, we apply the EVC model to study the data of hadronic
$v_{2}$ in heavy-ion collisions at RHIC energies. Here, we focus
on proton, $\Lambda$, $\Xi$, $\Omega$, and $\phi$. These hadrons
can be properly explained by constituent quark model with constituent
masses $m_{u}=m_{d}\approx0.3-0.33$ GeV and $m_{s}\approx0.5-0.55$
GeV. Therefore, their production can be suitably described by EVC
of constituent quarks at hadronization. However, pion and kaon, because
of their significantly small masses, can not be directly described
in the same way. We discuss their production in \ref{app_v2_piK}.

Using Eqs. (\ref{eq:Mv2_final_short}) and (\ref{eq:Bv2_final_short}),
we obtain
\begin{align}
v_{2,\Omega}\left(p_{T}\right) & =3v_{2,s}\left(p_{T}/3\right),\\
v_{2,p}\left(p_{T}\right) & =3v_{2,u}\left(p_{T}/3\right),\\
v_{2,\phi}\left(p_{T}\right) & =v_{2,s}\left(p_{T}/2\right)+v_{2,\bar{s}}\left(p_{T}/2\right),\label{eq:phiv2_ss}\\
v_{2,\Lambda}\left(p_{T}\right) & =2v_{2,u}\left(\frac{1}{2+r}p_{T}\right)+v_{2,s}\left(\frac{r}{2+r}p_{T}\right),\\
v_{2,\Xi}\left(p_{T}\right) & =v_{2,u}\left(\frac{1}{1+2r}p_{T}\right)+2v_{2,s}\left(\frac{r}{1+2r}p_{T}\right)
\end{align}
with the factor $r=m_{s}/m_{u}$=1.667. Here, we use $v_{2,u}=v_{2,d}$.

We can reversely obtain $v_{2}$ of $u$ quarks from proton or that
from hyperons
\begin{align}
v_{2,u}\left(p_{T}\right) & =\frac{1}{3}v_{2,p}\left(3p_{T}\right),\label{eq:v2up}\\
v_{2,u}\left(p_{T}\right) & =\frac{1}{3}\left[2v_{2,\Lambda}\left((2+r)p_{T}\right)-v_{2,\Xi}\left((1+2r)p_{T}\right)\right].\label{eq:v2uLamxi}
\end{align}
We can obtain $v_{2}$ of $s$ quarks from $\phi$ or hyperons
\begin{align}
v_{2,s}\left(p_{T}\right) & =\frac{1}{3}v_{2,\Omega}\left(3p_{T}\right),\label{eq:v2sOmg}\\
v_{2,s}\left(p_{T}\right) & =\frac{1}{2}v_{2,\phi}\left(2p_{T}\right),\label{eq:v2sphi}\\
v_{2,s}\left(p_{T}\right) & =\frac{1}{3}\left[2v_{2,\Xi}\left(\frac{1+2r}{r}p_{T}\right)-v_{2,\Lambda}\left(\frac{2+r}{r}p_{T}\right)\right].\label{eq:v2sLamXi}
\end{align}
Here, we use $v_{2,s}=v_{2,\bar{s}}$ in Eq. (\ref{eq:v2sphi}).

\begin{figure}[h]
\begin{centering}
\includegraphics[scale=0.4]{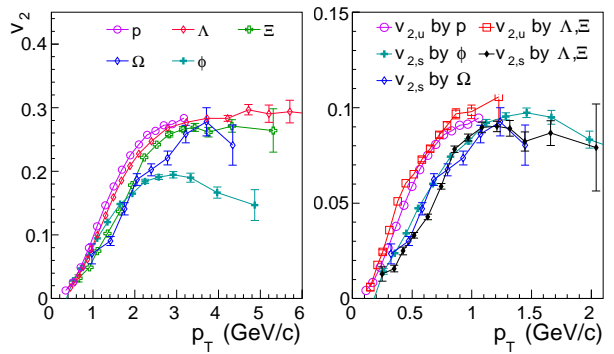}
    \caption{(a) Data for $v_{2}$ of identified hadrons at midrapidity in Au+Au collisions at $\sqrt{s_{NN}}=$200 GeV for 30-80\% centrality \cite{Adamczyk:2015ukd}; (b) $v_{2,u}$$\left(p_{T}\right)$ and $v_{2,s}\left(p_{T}\right)$
extracted from these hadrons. }
\label{fig1}
\par\end{centering}
\end{figure}

\begin{figure*}[bht]
\centering{}\includegraphics[scale=0.8]{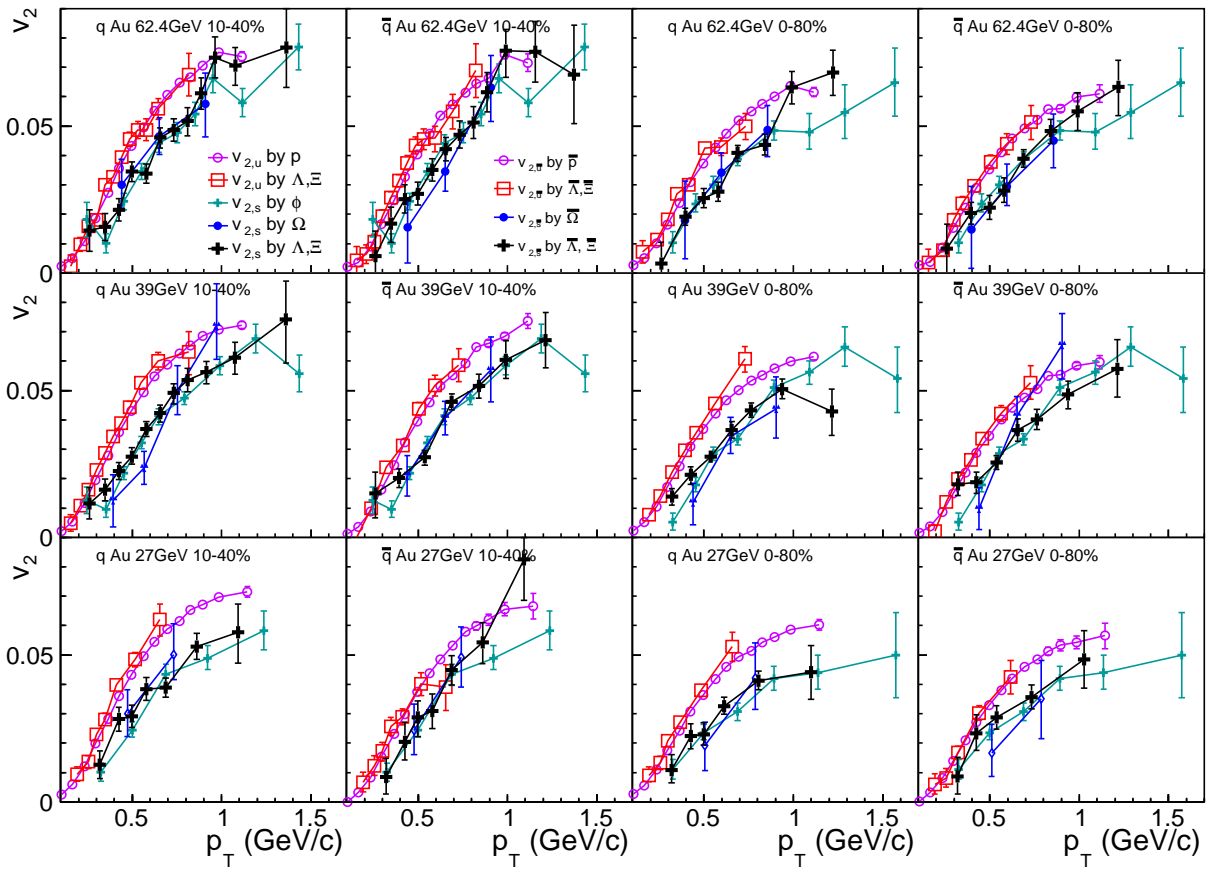}\caption{$v_{2,u}$$\left(p_{T}\right)$ and $v_{2,s}\left(p_{T}\right)$ extracted
from data of identified hadrons in Au+Au collisions in different centralities
and at different collision energies \cite{Adamczyk:2015fum}. }
\label{fig2}
\end{figure*}

As an example, we first test Eqs. (\ref{eq:v2up}-\ref{eq:v2sLamXi})
by experimental data of proton, $\Lambda$ ($\Lambda+\bar{\Lambda})$,
$\Xi$ ($\Xi^{-}+\bar{\Xi}^{+})$, $\Omega$ ($\Omega^{-}+\bar{\Omega}^{+}$),
and $\phi$ in Au+Au collisions at $\sqrt{s_{NN}}=$200 GeV for 30-80\%
centrality \cite{Adamczyk:2015ukd}. \ins{Data of proton, $\Lambda$, and $\Xi$ usually contain the decay contribution of heavier resonances. Because of decay kinematics, the influence of resonance decays on $v_2$ of these baryons is generally small. Therefore, we neglect this influence and  directly apply Eqs.
(\ref{eq:v2up}-\ref{eq:v2sLamXi}) to these hadrons.} We
also neglect the possible rescattering effect in hadronic stage for
the moment until we find its necessity in following analysis such
as in $\phi$ study at LHC energies in next section \ref{sec:v2_LHC}.

Panel (a) in Fig. \ref{fig1} shows the original data for $v_{2}$
of these hadrons which are different for different hadron species.
Panel (b) shows the extracted $v_{2,u}$$\left(p_{T}\right)$ and
$v_{2,s}\left(p_{T}\right)$ according to Eqs. (\ref{eq:v2up}-\ref{eq:v2sLamXi}).
We see that $v_{2,s}\left(p_{T}\right)$ extracted from $\Omega$
using Eq. (\ref{eq:v2sOmg}) is very close to that from $\phi$ using
Eq. (\ref{eq:v2sphi}) and is also very close to that from $\Lambda$
and $\Xi$ data using Eq. (\ref{eq:v2sLamXi}). For $u$ quarks, we
see that $v_{2,u}$$\left(p_{T}\right)$ extracted from proton data
using Eq. (\ref{eq:v2up}) is very close to that from $\Lambda$ and
$\Xi$ data using Eq.(\ref{eq:v2uLamxi}). Therefore, $v_{2}$ data
of these hadrons can be reasonably attributed to a common $v_{2,u}\left(p_{T}\right)$
and a common $v_{2,s}\left(p_{T}\right)$ at hadronization within
the experimental uncertainties. In addition, we see that the extracted
$v_{2,u}\left(p_{T}\right)$ is obviously larger than the extracted
$v_{2,s}\left(p_{T}\right)$ in the available range $p_{T}\lesssim1.2$
GeV/c, suggesting a flavor hierarchy property at quark level.

Furthermore, in Fig. \ref{fig2}, we carry out a systematic analysis
for STAR BES data in Au+Au collisions at $\sqrt{s_{NN}}=$ 62.4, 39
and 27 GeV \cite{Adamczyk:2015fum}. Because experimental data at
lower collision energies cover smaller $p_{T}$ range and have relatively
poor statistics, their results are not shown in this paper. Here,
data of hadrons and those of anti-hadrons are separately analyzed
to obtain $v_{2,u}\left(p_{T}\right)$ and $v_{2,\bar{u}}\left(p_{T}\right)$.
 In the figure, $v_{2,s}\left(p_{T}\right)$ obtained by $\phi$
using Eq. (\ref{eq:v2sphi}) is compared with those obtained from
baryons and also with those from anti-baryons. At these three collision
energies each with two centralities, we see that $v_{2,u}\left(p_{T}\right)$
obtained from $p$ and that from $\Lambda$ and $\Xi$ are consistent
with each other. The same case is for $v_{2,\bar{u}}\left(p_{T}\right)$.
Compared with $v_{2,u}\left(p_{T}\right)$ and $v_{2,\bar{u}}\left(p_{T}\right)$
data, $v_{2,s}\left(p_{T}\right)$ obtained from different strange
hadrons are limited by finite statistics but are also close to each
other.

\section{Quark $v_{2}$ at LHC and $\phi$ specificity \label{sec:v2_LHC}}

\begin{figure}
\centering{}\includegraphics[scale=0.38]{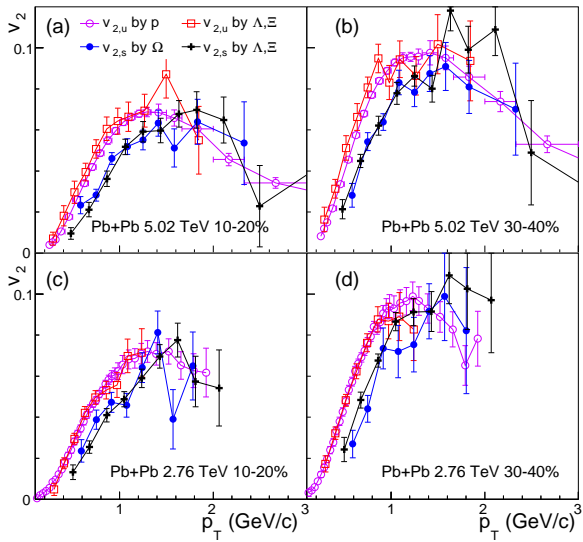}\caption{$v_{2,u}$$\left(p_{T}\right)$ and $v_{2,s}\left(p_{T}\right)$ extracted
from experimental data for $v_{2}$ of identified hadrons at midrapidity
in Pb+Pb collisions at LHC energies \cite{Acharya:2018zuq,Zhu:2018zqo}. }
\label{fig3}
\end{figure}

\begin{figure*}[t]
\centering{}\includegraphics[scale=0.75]{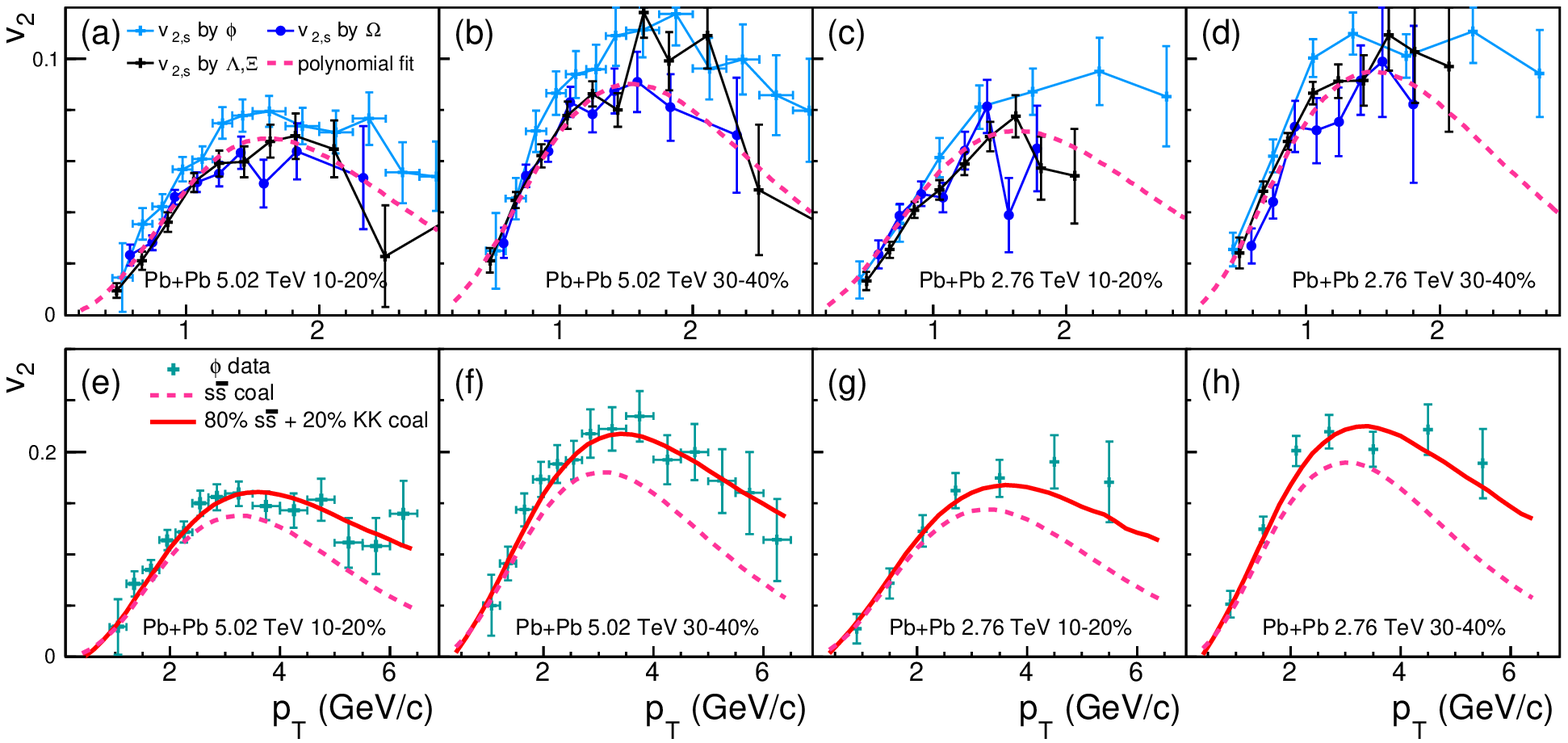}\caption{Top panels: $v_{2,s}\left(p_{T}\right)$ extracted from data of strange
hadrons \cite{Adamczyk:2015fum,Abelev:2014pua,Zhu:2018zqo}.The dashed
lines are polynomial fits of $v_{2,s}\left(p_{T}\right)$ from strange
baryons. Bottom panels: The $v_{2}$ of $\phi$. Symbols are data
of $\phi$ \cite{Adamczyk:2015fum,Abelev:2014pua,Zhu:2018zqo}. The
dashed lines, marked as $s\bar{s}$ coal, are results of strange quark
antiquark combination. The solid lines are results considering the
mixture of quark combination and two-kaon coalescence. See text for
details. }
\label{fig4}
\end{figure*}

We further study the property of $v_{2}$ under EVC by experimental
data in Pb+Pb collisions at LHC energies. In Fig. \ref{fig3}, we
present $v_{2,u}$$\left(p_{T}\right)$ and $v_{2,s}\left(p_{T}\right)$
extracted from midrapidity data of proton, $\Lambda,$ $\Xi$ and
$\Omega$ in Pb+Pb collisions at $\sqrt{s_{NN}}=$ 2.76 and 5.02 TeV
\cite{Acharya:2018zuq,Zhu:2018zqo}. We see that $v_{2,u}$$\left(p_{T}\right)$
extracted from proton and that from $\Lambda$ and $\Xi$ are very
consistent. $v_{2,s}\left(p_{T}\right)$ extracted from $\Omega$
and that from $\Lambda$ and $\Xi$ are also very close to each other.

We further consider the $\phi$ data \cite{Adamczyk:2015fum,Abelev:2014pua,Zhu:2018zqo}
to extract $v_{2,s}\left(p_{T}\right)$ and compare with those obtained
from strange baryons. The results are shown in top panels in Fig.
\ref{fig4}. Surprisingly, even though statistical uncertainties are
relatively large, we see that the $v_{2,s}\left(p_{T}\right)$ extracted
from $\phi$ seems to be higher than those from strange baryons to
a certain extent as $p_{T,s}\gtrsim1$ GeV/c. This is different from
the case at RHIC energies in Figs. \ref{fig1} and \ref{fig2}.

Results in top panels in Fig. \ref{fig4} imply that $\phi$ may receive
the contribution of other production channels in heavy-ion collisions
at LHC energies. Here, we consider a possible contribution, that is,
two-kaon coalescence $KK\to\phi$ in hadron scattering stage \cite{Baltz:1995tv,Alt:2008iv,Sun:2011kj}.
In this case, the distribution of final-state $\phi$ has two contributions
\begin{equation}
f_{\phi}^{\left(final\right)}\left(p_{T},\varphi\right)=f_{\phi,s\bar{s}}\left(p_{T},\varphi\right)+f_{\phi,KK}\left(p_{T},\varphi\right).
\end{equation}
The elliptic flow is
\begin{align}
v_{2,\phi}^{\left(final\right)} & \left(p_{T}\right)=\left(1-z\right)v_{2,s\bar{s}}\left(p_{T}\right)+z\,v_{2,KK}\left(p_{T}\right)
\end{align}
with the fraction

\begin{equation}
z=\frac{f_{\phi,KK}\left(p_{T}\right)}{f_{\phi,s\bar{s}}\left(p_{T}\right)+f_{\phi,KK}\left(p_{T}\right)}.
\end{equation}
 Using the relation Eq. (\ref{eq:phiv2_ss}) for $s\bar{s}$ combination
and the similar one for the coalescence of two kaons also with equal
velocity (because the mass $m_{\phi}\approx2m_{K}$ ), the elliptic
flow of $\phi$ after considering the possible two-kaon coalescence
has
\begin{equation}
v_{2,\phi}^{\left(final\right)}\left(p_{T}\right)\approx2\left[\left(1-z\right)v_{2,s}\left(\frac{p_{T}}{2}\right)+z\,v_{2,K}\left(\frac{p_{T}}{2}\right)\right].\label{eq:v2_phi_final}
\end{equation}
Here, we have taken $v_{2,\bar{s}}=v_{2,s}$ at LHC energies.

In bottom panels in Fig. \ref{fig4}, we calculate the elliptic flow
of $\phi$ and compare with experimental data \cite{Adamczyk:2015fum,Abelev:2014pua,Zhu:2018zqo}.
We firstly calculate the elliptic flow of pure $s\bar{s}$ combination.
The results, marked as $s\bar{s}$ coal, are shown as the dashed lines
in bottom panels in Fig. \ref{fig4}. The actual $v_{2,s}\left(p_{T}\right)$
at hadronization is identified as that by fitting data of strange
baryons, see the dashed lines in top panels in Fig. \ref{fig4}. We
see that pure $s\bar{s}$ combination can describe $\phi$ data in
low $p_{T}$ range ($p_{T}\lesssim2.5$ GeV/c) but under-estimates
the $v_{2}$ of $\phi$ at intermediate $p_{T}$ ( $p_{T}\gtrsim2.5$
GeV/c). We then consider the contribution of two-kaon coalescence
in a simple case that a $p_{T}$-independent $z$ is taken. Using
data of elliptic flow for kaons \cite{Abelev:2014pua,Acharya:2018zuq},
we calculate the elliptic flow of final-state $\phi$ by Eq. (\ref{eq:v2_phi_final})
and compare with the data. We find that data of $\phi$ at $p_{T}\gtrsim2.5$
GeV/c at two LHC energies can be roughly described by $z=0.2$, see
solid lines in bottom panels in Fig. \ref{fig4}. This implies that
there is about 20\% of $\phi$ with $p_{T}\gtrsim2.5$ GeV/c coming
from two-kaon coalescence in the hadronic stage. We note that, compared
with pure $s\bar{s}$ combination, two-kaon coalescence does not significantly
increase elliptic flow of $\phi$ in low $p_{T}$ range ($p_{T}\lesssim2$
GeV/c) because here $v_{2}$ of participant kaons is small as $p_{T,K}=p_{T}/2\lesssim1$
GeV/c. Therefore, we emphasize that $v_{2}$ data of $\phi$ in the
low $p_{T}$ range do not necessarily contain the contribution of
two-kaon coalescence. In addition, we note that two-kaon coalescence
will influence slightly the $p_{T}$ distribution function of $\phi$
and thus will slightly influence the quark number scaling property
for $p_{T}$ spectra of $\Omega$ and $\phi$ \cite{Song:2017gcz,Zhang:2018vyr,Song:2019sez}.
This influence is discussed in \ref{sec:appA}.

\section{Charm quark $v_{2}$ from $D$ mesons\label{sec:v2_charm}}

The EVC can be applied not only to light-flavor quarks but also to
heavy-flavor quarks at hadronization \cite{Lin:2003jy}. In \cite{Li:2017zuj,Song:2018tpv,Wang:2019fcg},
we show the EVC of charm quarks and soft light-flavor quarks provides
good description for $p_{T}$ spectra of single-charm hadrons in high
energy collisions. Applying EVC to elliptic flows of $D^{0,\pm}$
mesons, we obtain
\begin{align}
v_{2,D}\left(p_{T}\right) & =v_{2,u}\left(\frac{1}{1+r_{cu}}p_{T}\right)+v_{2,c}\left(\frac{r_{cu}}{1+r_{cu}}p_{T}\right)
\end{align}
with $r_{cu}=m_{c}/m_{u}=5$. Since we have obtained $v_{2,u}$ in
the previous sections, $v_{2}$ of charm quarks can be reversely extracted
by the data of $D$ mesons,
\begin{equation}
v_{2,c}\left(p_{T}\right)=v_{2,D}\left(\frac{1+r_{cu}}{r_{cu}}p_{T}\right)-v_{2,u}\left(\frac{1}{r_{cu}}p_{T}\right).
    \label{v2_c_by_D0}
\end{equation}
\ins{We note that this extraction is only valid in the low $p_T$ range where the combination dominates $D$ meson production. In previous studies \cite{Li:2017zuj,Song:2018tpv}, we found that experimental data for $p_T$ spectra of single-charmed hadrons in the range $p_T \lesssim 8$ GeV/c in pp and pPb collisions at LHC energies are well described by the EVC model. Therefore, experimental data of $v_2$ for $D$ mesons with $p_T\lesssim 8$ GeV/c can be used to extract $v_2$ of charm quarks with $p_T\lesssim 6$ GeV/c in EVC model. At larger $p_T$, fragmentation mechanism becomes important \cite{Oh:2009zj,Cao:2015hia} and Eq.~(\ref{v2_c_by_D0}) is no longer valid.}

\begin{figure}[h]
\centering{}\includegraphics[scale=0.4]{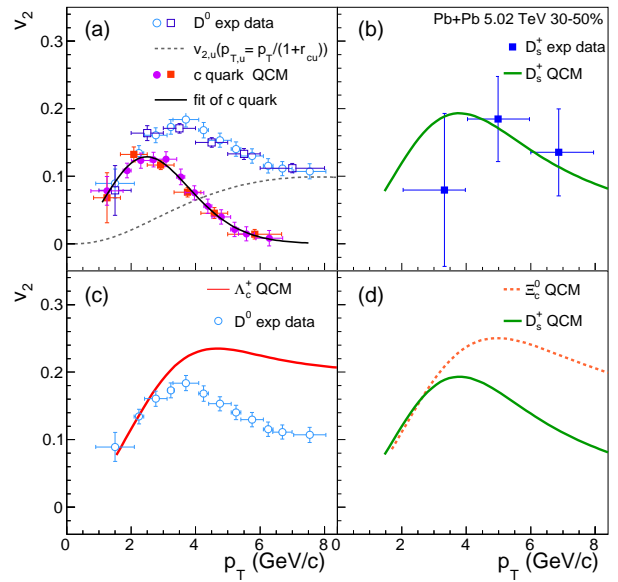}\caption{(a) Charm quark $v_{2}$ extracted from $D^{0}$ meson in Pb+Pb collisions
at $\sqrt{s_{NN}}=$ 5.02 TeV for 30-50\% centrality, and (b-d) the
predictions for $D_{s}^{+}$, $\Lambda_{c}^{+}$ and $\Xi_{c}^{0}$.
Data of $D^{0}$ and $D_{s}^{+}$ are taken from Ref. \cite{Sirunyan:2017plt,CMS:2019rjv,Vermunt:2019ecg}. }
\label{fig5}
\end{figure}

In Fig. \ref{fig5}(a), we show the extracted $v_{2,c}\left(p_{T}\right)$
in Pb+Pb collisions at $\sqrt{s_{NN}}=$ 5.02 TeV for 30-50\% centrality.
Solid circles and squares are results for $v_{2,c}\left(p_{T}\right)$
extracted from latest preliminary and previously published data of
$D^{0}$ \cite{Sirunyan:2017plt,CMS:2019rjv}, respectively. Data
of $D^{0}$ are also presented in the figure as open circles and squares,
respectively.
The contribution of up quarks to the $v_{2}$ of $D^{0}$
at different $p_{T}$ is shown as the dashed line. Under EVC, $v_{2}$
of $D^{0}$ at a specific $p_{T}$ absorbs the contribution of $u$
quark at a much smaller momentum $p_{T}/(1+r_{cu})$. Therefore, $v_{2}$
of $D^{0}$ in the low $p_{T}$ range ($p_{T}\lesssim3$ GeV/c) only
receives the small contribution of $u$ quark with $p_{T,u}\lesssim0.5$
GeV/c, see Fig. \ref{fig3}. However, $v_{2}$ of $D^{0}$ in
\ins{$5\lesssim p_{T}\lesssim 8$}
GeV/c contains the large contribution of $u$ quark with $p_{T,u}\gtrsim0.9$
GeV/c which is about 0.1 reading from Fig. \ref{fig3}. Subtracting
the $u$ quark contribution from $D^{0}$, $v_{2}$ of charm quarks
is obtained as the solid circles and squares in Fig. \ref{fig5} (a).
A smooth fit of these discrete points of charm quarks is shown as
the solid line. We see that the $v_{2}$ of charm quarks is close
to that of $D^{0}$ as $p_{T}\lesssim2$ GeV/c and is smaller
than the latter as \ins{$3\lesssim p_{T}\lesssim 6$}  GeV/c.

As $v_{2,c}$, $v_{2,u}$ and $v_{2,s}$ are obtained, we can predict
$v_{2}$ of $D_{s}^{+}$, $\Lambda_{c}^{+}$ and $\Xi_{c}^{0}$,
\begin{align}
v_{2,D_{s}}\left(p_{T}\right) & =v_{2,s}\left(\frac{1}{1+r_{cs}}p_{T}\right)+v_{2,c}\left(\frac{r_{cs}}{1+r_{cs}}p_{T}\right),\label{eq:v2Ds}
\end{align}
\begin{align}
v_{2,\Lambda_{c}}\left(p_{T}\right) & =2v_{2,u}\left(\frac{1}{2+r_{cu}}p_{T}\right)+v_{2,c}\left(\frac{r_{cu}}{2+r_{cu}}p_{T}\right),\label{eq:v2LamC}
\end{align}
\begin{align}
v_{2,\Xi_{c}}\left(p_{T}\right) & =v_{2,u}\left(\frac{1}{1+r+r_{cu}}p_{T}\right)+v_{2,s}\left(\frac{r}{1+r+r_{cu}}p_{T}\right),\nonumber \\
 & +v_{2,c}\left(\frac{r_{cu}}{1+r+r_{cu}}p_{T}\right)\label{eq:v2Xc}
\end{align}
with $r_{cs}=m_{c}/m_{s}=3$. Here, we neglect the statistical uncertainties
of the extracted datum points for $v_{2}$ of quarks and use their
smooth fits, i.e., the solid line for $v_{2,c}\left(p_{T}\right)$
in Fig. \ref{fig5} (a) and the dashed line for $v_{2,s}\left(p_{T}\right)$
in Fig \ref{fig4}(b), to calculate $v_{2}$ of $D_{s}^{+}$, $\Lambda_{c}^{+}$
and $\Xi_{c}^{0}$ by Eqs. (\ref{eq:v2Ds})-(\ref{eq:v2Xc}). Results
are shown in Fig. \ref{fig5} (b)-(d) as different types of lines.

Result of $D_{s}^{+}$ is compared with the preliminary data of ALICE
collaboration \cite{Vermunt:2019ecg}. Results of $\Lambda_{c}^{+}$
and $\Xi_{c}^{0}$ are compared with those of $D_{0}$ and $D_{s}^{+}$.
We see that $v_{2}$ of charmed baryons are close to those of charmed
mesons as $p_{T}\lesssim$ 3 GeV/c because the contribution of light-flavor
quarks is very small there. As $p_{T}\gtrsim$ 4 GeV/c, we see a significant
enhancement for $v_{2}$ of charmed baryons compared with those of
charmed mesons. This is because single-charm baryons absorb the $v_{2}$
of two light-flavor quarks at hadronization, see Eqs. (\ref{eq:v2LamC})
and (\ref{eq:v2Xc}), and the contribution of light-flavor quarks
becomes large as $p_{T}\gtrsim$ 4 GeV/c , e.g., see the dashed line
in Fig. \ref{fig5} (a) for the case of charmed mesons.

\begin{figure}[h]
\centering{}\includegraphics[scale=0.4]{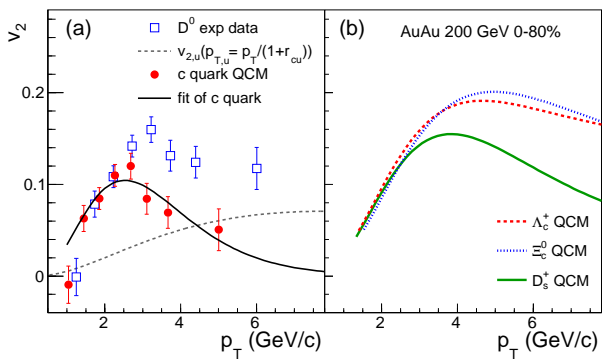}\caption{Panel (a) Charm quark $v_{2}$ extracted from midrapidity data of
$D^{0}$ in Au+Au collisions at $\sqrt{s_{NN}}=$ 200 GeV for 0-80\%
centrality \cite{Singha:2018cdj}, and (b) predictions for $D_{s}^{+}$,
$\Lambda_{c}^{+}$ and $\Xi_{c}^{0}$ in QCM. }
\label{fig6}
\end{figure}

In Fig. \ref{fig6} (a), we also study $v_{2}$ of charm quarks in
Au+Au collisions at $\sqrt{s_{NN}}=$200 GeV for 0-80\% centrality.
The result is similar to that in Pb+Pb collisions at $\sqrt{s_{NN}}=$
5.02 TeV. In the range of $p_{T}\lesssim3$ GeV/c, charm quark dominates
the $v_{2}$ of $D^{0}$ meson while at intermediate $p_{T}$ the
$u$ quark contributes significantly to the $v_{2}$ of $D^{0}$ meson.
Using the smooth fit of discrete points of charm quark $v_{2}$ in
panel (a) and those of light-flavor quarks in the corresponding centrality
extracted from data of light-flavor hadrons \cite{Adamczyk:2015ukd},
we predict in panel (b) the $v_{2}$ of $D_{s}^{+}$ meson, $\Lambda_{c}^{+}$
and $\Xi_{c}^{0}$ baryons. We see that in low $p_{T}$ range $v_{2}$
of $D_{s}^{+}$ is close to those of $\Lambda_{c}^{+}$ and $\Xi_{c}^{0}$
and at intermediate $p_{T}$ it is smaller than the latter. We also
see that $v_{2}$ of $\Lambda_{c}^{+}$ is slightly smaller than that
of $\Xi_{c}^{0}$ in the range $4\lesssim p_{T}\lesssim6$ GeV/c,
which is the kinetic effect caused by the mass difference of $u$
and $s$ quarks in combination with charm quark.

\section{Properties for $v_{2}$ of quark\label{sec:quark_v2_property}}

In this section, we study the property for $v_{2}$ of up, strange
and charm quarks obtained in previous sections. We focus on two main
properties, i.e., flavor dependence and the quark-antiquark split,
which are discussed as follows.

\subsection{compare $v_{2}$ of $u$, $s$ and $c$ quarks }

\begin{figure}[h]
\centering{}\includegraphics[scale=0.42]{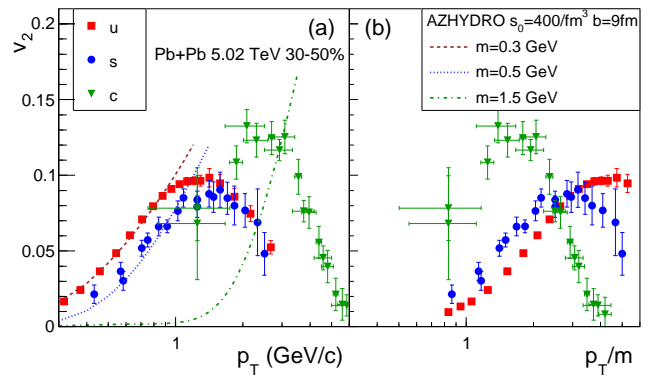}\caption{$v_{2}$ of up, strange and charm quarks in Pb+Pb collisions at $\sqrt{s_{NN}}=$
5.02 TeV for 30-50\% centrality as a function of $p_{T}$ (a) and
of $p_{T}/m$ (b). Lines are results of AZHYDRO code with initial
entropy density $s_{0}=400/fm^{3}$, impact parameter $b=9$ fm, and
hadronization temperature $T=0.165$ GeV. }
\label{fig7}
\end{figure}

As an example, in Fig. \ref{fig7} (a), we present $v_{2}$ of up,
strange and charm quarks in Pb+Pb collisions at $\sqrt{s_{NN}}=$
5.02 TeV for 30-50\% centrality. We see that $v_{2}$ of up and strange
quarks increase with $p_{T}$ as $p_{T}\lesssim$ 1.5 GeV/c and start
to decrease at larger $p_{T}$. In particular, we see that $v_{2}$
of up quarks is higher than that of strange quarks in the range $p_{T}\lesssim$
1.5 GeV/c. For all the studied collisions energies and collision centralities,
we always see this property. We have checked that if we replace $p_{T}$
by $\sqrt{p_{T}^{2}+m^{2}}-m$ in the horizontal axis, the split between
up and strange quarks becomes small but does not disappear. $v_{2}$
of charm quarks at small $p_{T}\lesssim$ 2 GeV/c has relatively large
uncertainty and is roughly consistent with those of up and strange
quarks. However, charm quark $v_{2}$ continues to increase and reaches
the peak value about 0.13 at $p_{T}\approx2.5$ GeV/c, which is obviously
higher than those of light-flavor quarks (about 0.09) at $p_{T}\approx1.5$
GeV/c.

For the increase of $v_{2}$ at small $p_{T}$ for up and strange
quarks, we can qualitatively understand it in general by the hydrodynamic
evolution of quark gluon plasma (QGP) \cite{Kolb:2000sd}. The $v_{2}$
for charm quarks is the result of the diffusion in QGP \cite{Rapp:2009my},
which is related to many evolution dynamics such as the large collective
flow, quench effects of the background medium and possible thermalization
of charm quarks \cite{Svetitsky:1987gq,Moore:2004tg,vanHees:2004gq,Rapp:2009my,He:2011qa,Cao:2013ita,Das:2015ana}.

\ins{To understand the flavor differenence of $v_2$ at small $p_T$, we apply the AZHYDRO code \cite{Kolb:2000sd} for 2+1-dimensional hydrodynamics to study the qualitative behavior for $v_2$ of different quark flavors under thermal equilibrium and hydrodynamic flow velocity field. We set the criterion of Cooper-Frye procedure by the fixed temperature. The temperature is taken as the hadronization temperature $T=0.165$ GeV. We change the ``freeze-out" particles as quarks in Cooper-Frye procedure and calculate the two-dimensional transverse momentum distributions of up, strange and charm quarks.}
The initial entropy density is set to be $s_{0}=400/fm^{3}$ and
impact parameter is set to be $b=9$ fm.
The inelastic $pp$ cross section is set to be 70 mb.
\ins{Calculation results for $v_2$ of quarks are shown as lines in Fig. \ref{fig7}(a).}
We see AZHYDRO simulations, the dashed and dotted lines in Fig. \ref{fig7}(a),
can well fit the extracted $v_{2}$ of up and strange quarks as $p_{T}\lesssim$
1 GeV/c. $v_{2}$ of charm quarks in the case of thermal equilibrium
is also shown as the dot-dashed line. It is below the extracted charm
$v_{2}$ at small $p_{T}$ and intersects the latter at $p_{T}\approx$
3 GeV/c.

Because masses of up, strange and charm quarks are quite different,
$p_{T}/m=\gamma v$ may be an alternative kinetic variable to reveal
the regularity for $v_{2}$ of three quark flavors. In panel
(b), we show quark $v_{2}$ as the function of $p_{T}/m$. Here, we
see a clear property relating to quark mass: as $p_{T}/m\lesssim2$
the quark with heavier mass has larger $v_{2}$ while the reverse
behavior appears as $p_{T}/m\gtrsim3$. We observe the same property
in Au+Au collisions at $\sqrt{s_{NN}}=200$ GeV for 0-80\% centrality.
This regularity of quark $v_{2}$ is interesting and is worthwhile
to be studied in the future work.

\subsection{$v_{2}$ split between quark and antiquark}

\begin{figure*}
\centering{}\includegraphics[scale=0.8]{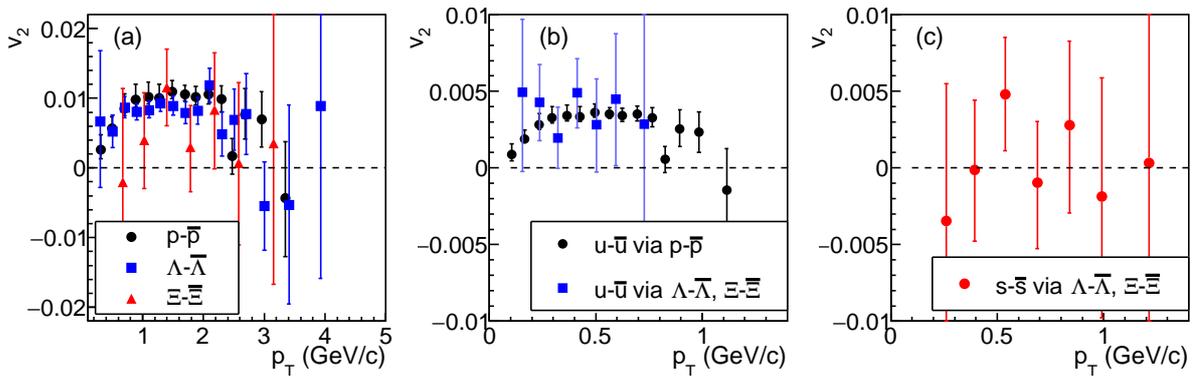}\caption{Panel(a): The difference in $v_{2}$ between hadrons and their anti-hadrons
in Au+Au collisions at $\sqrt{s_{NN}}=39$ GeV for 10-40\%
centrality. Data are from \cite{Adamczyk:2015fum}; panel (b) the
difference in $v_{2}$ between $u$ and $\bar{u}$; panel (c) that
between $s$ and $\bar{s}$.}
\label{fig8}
\end{figure*}

STAR experiments observed the elliptic flow split between hadrons
and their anti-hadrons at low collision energies \cite{Adamczyk:2015fum}.
In this paper, as shown in Fig. \ref{fig2}, we can apply the EVC
model to successfully decompose $v_{2}$ of hadrons and their anti-hadrons
into $v_{2}$ of quarks and antiquarks. Therefore, we can extract
the $v_{2}$ split between quark and antiquark. Here, we take the
data in Au+Au collisions at $\sqrt{s_{NN}}=39$ GeV for 10-40\%
centrality as an example.

In Fig. \ref{fig8} (a), we first present experimental data for the
difference in $v_{2}$ between baryons and their anti-hadrons. We
see that $v_{2,p}-v_{2,\bar{p}}$ show clearly positive values. $v_{2,\Lambda}-v_{2,\bar{\Lambda}}$
is also positive and is slightly smaller than $v_{2,p}-v_{2,\bar{p}}$.
Data of $v_{2,\Xi}-v_{2,\bar{\Xi}}$ have relatively poor statistics
and still show a positive tendency and smaller magnitude compared
to $v_{2,p}-v_{2,\bar{p}}$ and $v_{2,\Lambda}-v_{2,\bar{\Lambda}}$.
Data of $\Omega^{-}-\bar{\Omega}^{+}$ are not shown here because
of bad statistics.

Using $v_{2}$ of quarks and that of antiquarks obtained in Fig. \ref{fig2},
we calculate $v_{2,u}-v_{2,\bar{u}}$ and show results in Fig. \ref{fig8}
(b). We obtain a good agreement between results obtained from $p-\bar{p}$
via Eq. (\ref{eq:v2up}) and those from $\Lambda-\bar{\Lambda}$ and
$\Xi-\bar{\Xi}$ via Eq. (\ref{eq:v2uLamxi}). We see clearly the
positive $v_{2,u}-v_{2,\bar{u}}$ with weak $p_{T}$ dependence in
the low $p_{T}$ range ($p_{T}<1$ GeV/c).

Results of $v_{2,s}-v_{2,\bar{s}}$ from $\Lambda-\bar{\Lambda}$
and $\Xi-\bar{\Xi}$ channel are shown in Fig. \ref{fig8} (c). Because
of large statistical uncertainties, the $p_{T}$ dependence of $v_{2,s}-v_{2,\bar{s}}$
is not conclusive. For the overall sign of $v_{2,s}-v_{2,\bar{s}}$,
we can roughly estimate by averaging seven datum points and obtain
$0.0002\pm0.0029$, which might imply the equal $v_{2}$ for $s$
and $\bar{s}$ in Au+Au collisions at $\sqrt{s_{NN}}=39$
GeV. Results of $v_{2,s}-v_{2,\bar{s}}$ at lower collision energies
have poorer statistics and therefore we cannot draw further conclusion
at present.

\section{Summary and outlook \label{sec:summary}}

We applied a quark combination model (QCM) with equal-velocity combination
(EVC) approximation to study the elliptic flow $(v_{2}$) of identified
hadrons in relativistic heavy-ion collisions at $\sqrt{s_{NN}}=$
27$\sim$5020GeV. Under EVC, quarks contribute to the momentum of
the formed hadron by the fraction proportional to their constituent
masses. $v_{2}$ of hadron consisting of different constituent quarks
is therefore the sum of $v_{2}$ of quarks with different transverse
momenta. This is different from the number-of-constituent quark scaling
(NCQ) operation in experimental study of hadronic $v_{2}$ \cite{Adare:2006ti,Adamczyk:2015ukd,Abelev:2014pua,Acharya:2018zuq,Adamczyk:2013gw}.

Under EVC, we obtained simple formulas of reversely extracting $v_{2}$
of quarks and antiquarks from the experimental data of identified
hadrons. By the combination of data for $\Lambda$ and $\Xi$, we
obtained $v_{2}$ of up/down quarks which is consistent with that
from data of proton; we also obtained $v_{2}$ of strange quarks which
is consistent with that from data of $\Omega$. At RHIC energies,
$v_{2}$ of strange quarks extracted from hyperons is also consistent
with that from $\phi$ meson. This means that $v_{2}$ of these light-flavor
hadrons have a common quark-level source at hadronization. At LHC
energies, however, $v_{2}$ of strange quarks extracted from hyperons
is somewhat lower than that from $\phi$. This indicates the possible
contribution of two-kaon coalescence to $\phi$ production at LHC
energies.

Using results for $v_{2}$ of light flavor quarks, we further extracted
$v_{2}$ of charm quarks from the data of $D^{0}$ meson in Pb+Pb
collisions at $\sqrt{s_{NN}}=$ 5.02 TeV and that in Au+Au collisions
at 200 GeV. By comparing $v_{2}$ of charm quark and that of $D^{0}$
meson, we found that $v_{2}$ of $D^{0}$ meson at low $p_{T}$ ($p_{T}\lesssim$
3 GeV/c) is dominated by that of charm quarks but at intermediate
$p_{T}$ ($3\lesssim p_{T}\lesssim8$ GeV/c) is significantly contributed
by $v_{2}$ of light-flavor quarks. We predicted $v_{2}$ of $D_{s}^{+}$
meson, $\Lambda_{c}^{+}$ and $\Xi_{c}^{0}$ baryons. We found that
$v_{2}$ of charmed baryons is significantly enhanced at intermediate
$p_{T}$ ($3\lesssim p_{T}\lesssim8$ GeV/c), compared to those of
$D$ mesons, which is due to the double contribution of light-flavor
quarks.

We finally studied the properties of the extracted $v_{2}$ of quarks
and antiquarks at hadronization. We first compared $v_{2}$ of up,
strange and charm quarks. We found that $v_{2}$ of up quarks is always
higher than that of strange quarks at low $p_{T}$ ($p_{T}\lesssim$
1 GeV/c) at all studied collision energies. Such a mass hierarchy
of quark $v_{2}$ can be reasonably understood by hydrodynamics. $v_{2}$
of charm quark at small $p_{T}$ ($p_{T}\lesssim1.5$ GeV/c) is roughly
consistent with those of light-flavor quarks within the statistical
uncertainty. However, differing from light-flavor quarks, $v_{2}$
of charm quarks continues to increase with $p_{T}$ and reaches larger
value at $p_{T}\approx2.5$ GeV/c. Interestingly, by plotting quark
$v_{2}$ as the function of transverse velocity $p_{T}/m$, we found
a regularity relating to quark mass, i.e., as $p_{T}/m\lesssim2$
the quark with heavier mass has larger $v_{2}$ while as $p_{T}/m\gtrsim3$
reverse property holds. We further studied the difference in $v_{2}$
between quarks and antiquarks at low RHIC energies. We found that
$v_{2,u}-v_{2,\bar{u}}$ extracted from hyperons and anti-hyperons
coincides with that from proton and anti-proton. Results of $v_{2,s}-v_{2,\bar{s}}$
have large statistical uncertainties, and the average value of all
datum points implies the symmetry in $v_{2}$ between strange quarks
and strange antiquarks in Au+Au collisions at 39 GeV.

These results suggest that QCM with EVC is quite effective and self-consistent
in understanding $v_{2}$ of hadrons in heavy-ion collisions. In addition,
EVC mechanism provides \ins{an approximate method to obtain} $v_{2}$ of quarks and
antiquarks at hadronization, by which we can obtain deeper insights
into the information of partonic stage evolution in relativistic heavy-ion
collisions. \ins{Finally, we emphasize that EVC is an effective but simplifed mechanism.} Studies \ins{on the application range and limitation of this mechanism} are deserved with the help
of more precise experimental data in the future.

\section{Acknowledgments}
We gratefully acknowledge fruitful discussions with Z.T. Liang.  This work is supported in part by the National Natural Science Foundation of China under Grant Nos. 11975011, Shandong Province Natural Science Foundation Grant Nos. ZR2019YQ06 and ZR2019MA053, and Higher Educational Youth Innovation Science and Technology Program of Shandong Province (2019KJJ010). 

\bibliographystyle{apsrev4-1}
\bibliography{ref}

\begin{thebibliography}{59}%
\makeatletter
\providecommand \@ifxundefined [1]{%
 \@ifx{#1\undefined}
}%
\providecommand \@ifnum [1]{%
 \ifnum #1\expandafter \@firstoftwo
 \else \expandafter \@secondoftwo
 \fi
}%
\providecommand \@ifx [1]{%
 \ifx #1\expandafter \@firstoftwo
 \else \expandafter \@secondoftwo
 \fi
}%
\providecommand \natexlab [1]{#1}%
\providecommand \enquote  [1]{``#1''}%
\providecommand \bibnamefont  [1]{#1}%
\providecommand \bibfnamefont [1]{#1}%
\providecommand \citenamefont [1]{#1}%
\providecommand \href@noop [0]{\@secondoftwo}%
\providecommand \href [0]{\begingroup \@sanitize@url \@href}%
\providecommand \@href[1]{\@@startlink{#1}\@@href}%
\providecommand \@@href[1]{\endgroup#1\@@endlink}%
\providecommand \@sanitize@url [0]{\catcode `\\12\catcode `\$12\catcode
  `\&12\catcode `\#12\catcode `\^12\catcode `\_12\catcode `\%12\relax}%
\providecommand \@@startlink[1]{}%
\providecommand \@@endlink[0]{}%
\providecommand \url  [0]{\begingroup\@sanitize@url \@url }%
\providecommand \@url [1]{\endgroup\@href {#1}{\urlprefix }}%
\providecommand \urlprefix  [0]{URL }%
\providecommand \Eprint [0]{\href }%
\providecommand \doibase [0]{http://dx.doi.org/}%
\providecommand \selectlanguage [0]{\@gobble}%
\providecommand \bibinfo  [0]{\@secondoftwo}%
\providecommand \bibfield  [0]{\@secondoftwo}%
\providecommand \translation [1]{[#1]}%
\providecommand \BibitemOpen [0]{}%
\providecommand \bibitemStop [0]{}%
\providecommand \bibitemNoStop [0]{.\EOS\space}%
\providecommand \EOS [0]{\spacefactor3000\relax}%
\providecommand \BibitemShut  [1]{\csname bibitem#1\endcsname}%
\let\auto@bib@innerbib\@empty
\bibitem [{\citenamefont {Ollitrault}(1992)}]{Ollitrault:1992bk}%
  \BibitemOpen
  \bibfield  {author} {\bibinfo {author} {\bibfnamefont {J.-Y.}\ \bibnamefont
  {Ollitrault}},\ }\href {\doibase 10.1103/PhysRevD.46.229} {\bibfield
  {journal} {\bibinfo  {journal} {Phys. Rev.}\ }\textbf {\bibinfo {volume}
  {D46}},\ \bibinfo {pages} {229} (\bibinfo {year} {1992})}\BibitemShut
  {NoStop}%
\bibitem [{\citenamefont {Sorge}(1999)}]{Sorge:1998mk}%
  \BibitemOpen
  \bibfield  {author} {\bibinfo {author} {\bibfnamefont {H.}~\bibnamefont
  {Sorge}},\ }\href {\doibase 10.1103/PhysRevLett.82.2048} {\bibfield
  {journal} {\bibinfo  {journal} {Phys. Rev. Lett.}\ }\textbf {\bibinfo
  {volume} {82}},\ \bibinfo {pages} {2048} (\bibinfo {year} {1999})},\ \Eprint
  {http://arxiv.org/abs/nucl-th/9812057} {arXiv:nucl-th/9812057 [nucl-th]}
  \BibitemShut {NoStop}%
\bibitem [{\citenamefont {Teaney}\ \emph {et~al.}(2001)\citenamefont {Teaney},
  \citenamefont {Lauret},\ and\ \citenamefont {Shuryak}}]{Teaney:2000cw}%
  \BibitemOpen
  \bibfield  {author} {\bibinfo {author} {\bibfnamefont {D.}~\bibnamefont
  {Teaney}}, \bibinfo {author} {\bibfnamefont {J.}~\bibnamefont {Lauret}}, \
  and\ \bibinfo {author} {\bibfnamefont {E.~V.}\ \bibnamefont {Shuryak}},\
  }\href {\doibase 10.1103/PhysRevLett.86.4783} {\bibfield  {journal} {\bibinfo
   {journal} {Phys. Rev. Lett.}\ }\textbf {\bibinfo {volume} {86}},\ \bibinfo
  {pages} {4783} (\bibinfo {year} {2001})},\ \Eprint
  {http://arxiv.org/abs/nucl-th/0011058} {arXiv:nucl-th/0011058 [nucl-th]}
  \BibitemShut {NoStop}%
\bibitem [{\citenamefont {Huovinen}\ \emph {et~al.}(2001)\citenamefont
  {Huovinen}, \citenamefont {Kolb}, \citenamefont {Heinz}, \citenamefont
  {Ruuskanen},\ and\ \citenamefont {Voloshin}}]{Huovinen:2001cy}%
  \BibitemOpen
  \bibfield  {author} {\bibinfo {author} {\bibfnamefont {P.}~\bibnamefont
  {Huovinen}}, \bibinfo {author} {\bibfnamefont {P.~F.}\ \bibnamefont {Kolb}},
  \bibinfo {author} {\bibfnamefont {U.~W.}\ \bibnamefont {Heinz}}, \bibinfo
  {author} {\bibfnamefont {P.~V.}\ \bibnamefont {Ruuskanen}}, \ and\ \bibinfo
  {author} {\bibfnamefont {S.~A.}\ \bibnamefont {Voloshin}},\ }\href {\doibase
  10.1016/S0370-2693(01)00219-2} {\bibfield  {journal} {\bibinfo  {journal}
  {Phys. Lett.}\ }\textbf {\bibinfo {volume} {B503}},\ \bibinfo {pages} {58}
  (\bibinfo {year} {2001})},\ \Eprint {http://arxiv.org/abs/hep-ph/0101136}
  {arXiv:hep-ph/0101136 [hep-ph]} \BibitemShut {NoStop}%
\bibitem [{\citenamefont {Adare}\ \emph {et~al.}(2007)\citenamefont {Adare}
  \emph {et~al.}}]{Adare:2006ti}%
  \BibitemOpen
  \bibfield  {author} {\bibinfo {author} {\bibfnamefont {A.}~\bibnamefont
  {Adare}} \emph {et~al.} (\bibinfo {collaboration} {PHENIX}),\ }\href
  {\doibase 10.1103/PhysRevLett.98.162301} {\bibfield  {journal} {\bibinfo
  {journal} {Phys. Rev. Lett.}\ }\textbf {\bibinfo {volume} {98}},\ \bibinfo
  {pages} {162301} (\bibinfo {year} {2007})},\ \Eprint
  {http://arxiv.org/abs/nucl-ex/0608033} {arXiv:nucl-ex/0608033 [nucl-ex]}
  \BibitemShut {NoStop}%
\bibitem [{\citenamefont {Adamczyk}\ \emph
  {et~al.}(2016{\natexlab{a}})\citenamefont {Adamczyk} \emph
  {et~al.}}]{Adamczyk:2015ukd}%
  \BibitemOpen
  \bibfield  {author} {\bibinfo {author} {\bibfnamefont {L.}~\bibnamefont
  {Adamczyk}} \emph {et~al.} (\bibinfo {collaboration} {STAR}),\ }\href
  {\doibase 10.1103/PhysRevLett.116.062301} {\bibfield  {journal} {\bibinfo
  {journal} {Phys. Rev. Lett.}\ }\textbf {\bibinfo {volume} {116}},\ \bibinfo
  {pages} {062301} (\bibinfo {year} {2016}{\natexlab{a}})},\ \Eprint
  {http://arxiv.org/abs/1507.05247} {arXiv:1507.05247 [nucl-ex]} \BibitemShut
  {NoStop}%
\bibitem [{\citenamefont {Abelev}\ \emph {et~al.}(2015)\citenamefont {Abelev}
  \emph {et~al.}}]{Abelev:2014pua}%
  \BibitemOpen
  \bibfield  {author} {\bibinfo {author} {\bibfnamefont {B.~B.}\ \bibnamefont
  {Abelev}} \emph {et~al.} (\bibinfo {collaboration} {ALICE}),\ }\href
  {\doibase 10.1007/JHEP06(2015)190} {\bibfield  {journal} {\bibinfo  {journal}
  {JHEP}\ }\textbf {\bibinfo {volume} {06}},\ \bibinfo {pages} {190} (\bibinfo
  {year} {2015})},\ \Eprint {http://arxiv.org/abs/1405.4632} {arXiv:1405.4632
  [nucl-ex]} \BibitemShut {NoStop}%
\bibitem [{\citenamefont {Acharya}\ \emph {et~al.}(2018)\citenamefont {Acharya}
  \emph {et~al.}}]{Acharya:2018zuq}%
  \BibitemOpen
  \bibfield  {author} {\bibinfo {author} {\bibfnamefont {S.}~\bibnamefont
  {Acharya}} \emph {et~al.} (\bibinfo {collaboration} {ALICE}),\ }\href
  {\doibase 10.1007/JHEP09(2018)006} {\bibfield  {journal} {\bibinfo  {journal}
  {JHEP}\ }\textbf {\bibinfo {volume} {09}},\ \bibinfo {pages} {006} (\bibinfo
  {year} {2018})},\ \Eprint {http://arxiv.org/abs/1805.04390} {arXiv:1805.04390
  [nucl-ex]} \BibitemShut {NoStop}%
\bibitem [{\citenamefont {Adamczyk}\ \emph {et~al.}(2013)\citenamefont
  {Adamczyk} \emph {et~al.}}]{Adamczyk:2013gw}%
  \BibitemOpen
  \bibfield  {author} {\bibinfo {author} {\bibfnamefont {L.}~\bibnamefont
  {Adamczyk}} \emph {et~al.} (\bibinfo {collaboration} {STAR}),\ }\href
  {\doibase 10.1103/PhysRevC.88.014902} {\bibfield  {journal} {\bibinfo
  {journal} {Phys. Rev.}\ }\textbf {\bibinfo {volume} {C88}},\ \bibinfo {pages}
  {014902} (\bibinfo {year} {2013})},\ \Eprint {http://arxiv.org/abs/1301.2348}
  {arXiv:1301.2348 [nucl-ex]} \BibitemShut {NoStop}%
\bibitem [{\citenamefont {Molnar}\ and\ \citenamefont
  {Voloshin}(2003)}]{Molnar:2003ff}%
  \BibitemOpen
  \bibfield  {author} {\bibinfo {author} {\bibfnamefont {D.}~\bibnamefont
  {Molnar}}\ and\ \bibinfo {author} {\bibfnamefont {S.~A.}\ \bibnamefont
  {Voloshin}},\ }\href {\doibase 10.1103/PhysRevLett.91.092301} {\bibfield
  {journal} {\bibinfo  {journal} {Phys. Rev. Lett.}\ }\textbf {\bibinfo
  {volume} {91}},\ \bibinfo {pages} {092301} (\bibinfo {year} {2003})},\
  \Eprint {http://arxiv.org/abs/nucl-th/0302014} {arXiv:nucl-th/0302014
  [nucl-th]} \BibitemShut {NoStop}%
\bibitem [{\citenamefont {Fries}\ \emph
  {et~al.}(2003{\natexlab{a}})\citenamefont {Fries}, \citenamefont {Muller},
  \citenamefont {Nonaka},\ and\ \citenamefont {Bass}}]{Fries:2003kq}%
  \BibitemOpen
  \bibfield  {author} {\bibinfo {author} {\bibfnamefont {R.}~\bibnamefont
  {Fries}}, \bibinfo {author} {\bibfnamefont {B.}~\bibnamefont {Muller}},
  \bibinfo {author} {\bibfnamefont {C.}~\bibnamefont {Nonaka}}, \ and\ \bibinfo
  {author} {\bibfnamefont {S.}~\bibnamefont {Bass}},\ }\href {\doibase
  10.1103/PhysRevC.68.044902} {\bibfield  {journal} {\bibinfo  {journal} {Phys.
  Rev. C}\ }\textbf {\bibinfo {volume} {68}},\ \bibinfo {pages} {044902}
  (\bibinfo {year} {2003}{\natexlab{a}})},\ \Eprint
  {http://arxiv.org/abs/nucl-th/0306027} {arXiv:nucl-th/0306027} \BibitemShut
  {NoStop}%
\bibitem [{\citenamefont {Fries}\ \emph
  {et~al.}(2003{\natexlab{b}})\citenamefont {Fries}, \citenamefont {Muller},
  \citenamefont {Nonaka},\ and\ \citenamefont {Bass}}]{Fries:2003vb}%
  \BibitemOpen
  \bibfield  {author} {\bibinfo {author} {\bibfnamefont {R.}~\bibnamefont
  {Fries}}, \bibinfo {author} {\bibfnamefont {B.}~\bibnamefont {Muller}},
  \bibinfo {author} {\bibfnamefont {C.}~\bibnamefont {Nonaka}}, \ and\ \bibinfo
  {author} {\bibfnamefont {S.}~\bibnamefont {Bass}},\ }\href {\doibase
  10.1103/PhysRevLett.90.202303} {\bibfield  {journal} {\bibinfo  {journal}
  {Phys. Rev. Lett.}\ }\textbf {\bibinfo {volume} {90}},\ \bibinfo {pages}
  {202303} (\bibinfo {year} {2003}{\natexlab{b}})},\ \Eprint
  {http://arxiv.org/abs/nucl-th/0301087} {arXiv:nucl-th/0301087} \BibitemShut
  {NoStop}%
\bibitem [{\citenamefont {Greco}\ \emph {et~al.}(2003)\citenamefont {Greco},
  \citenamefont {Ko},\ and\ \citenamefont {Levai}}]{Greco:2003xt}%
  \BibitemOpen
  \bibfield  {author} {\bibinfo {author} {\bibfnamefont {V.}~\bibnamefont
  {Greco}}, \bibinfo {author} {\bibfnamefont {C.}~\bibnamefont {Ko}}, \ and\
  \bibinfo {author} {\bibfnamefont {P.}~\bibnamefont {Levai}},\ }\href
  {\doibase 10.1103/PhysRevLett.90.202302} {\bibfield  {journal} {\bibinfo
  {journal} {Phys. Rev. Lett.}\ }\textbf {\bibinfo {volume} {90}},\ \bibinfo
  {pages} {202302} (\bibinfo {year} {2003})},\ \Eprint
  {http://arxiv.org/abs/nucl-th/0301093} {arXiv:nucl-th/0301093} \BibitemShut
  {NoStop}%
\bibitem [{\citenamefont {Kolb}\ \emph {et~al.}(2004)\citenamefont {Kolb},
  \citenamefont {Chen}, \citenamefont {Greco},\ and\ \citenamefont
  {Ko}}]{Kolb:2004gi}%
  \BibitemOpen
  \bibfield  {author} {\bibinfo {author} {\bibfnamefont {P.~F.}\ \bibnamefont
  {Kolb}}, \bibinfo {author} {\bibfnamefont {L.-W.}\ \bibnamefont {Chen}},
  \bibinfo {author} {\bibfnamefont {V.}~\bibnamefont {Greco}}, \ and\ \bibinfo
  {author} {\bibfnamefont {C.~M.}\ \bibnamefont {Ko}},\ }\href {\doibase
  10.1103/PhysRevC.69.051901} {\bibfield  {journal} {\bibinfo  {journal} {Phys.
  Rev. C}\ }\textbf {\bibinfo {volume} {69}},\ \bibinfo {pages} {051901}
  (\bibinfo {year} {2004})},\ \Eprint {http://arxiv.org/abs/nucl-th/0402049}
  {arXiv:nucl-th/0402049} \BibitemShut {NoStop}%
\bibitem [{\citenamefont {Minissale}\ \emph {et~al.}(2015)\citenamefont
  {Minissale}, \citenamefont {Scardina},\ and\ \citenamefont
  {Greco}}]{Minissale:2015zwa}%
  \BibitemOpen
  \bibfield  {author} {\bibinfo {author} {\bibfnamefont {V.}~\bibnamefont
  {Minissale}}, \bibinfo {author} {\bibfnamefont {F.}~\bibnamefont {Scardina}},
  \ and\ \bibinfo {author} {\bibfnamefont {V.}~\bibnamefont {Greco}},\ }\href
  {\doibase 10.1103/PhysRevC.92.054904} {\bibfield  {journal} {\bibinfo
  {journal} {Phys. Rev. C}\ }\textbf {\bibinfo {volume} {92}},\ \bibinfo
  {pages} {054904} (\bibinfo {year} {2015})},\ \Eprint
  {http://arxiv.org/abs/1502.06213} {arXiv:1502.06213 [nucl-th]} \BibitemShut
  {NoStop}%
\bibitem [{\citenamefont {Oh}\ \emph {et~al.}(2009)\citenamefont {Oh},
  \citenamefont {Ko}, \citenamefont {Lee},\ and\ \citenamefont
  {Yasui}}]{Oh:2009zj}%
  \BibitemOpen
  \bibfield  {author} {\bibinfo {author} {\bibfnamefont {Y.}~\bibnamefont
  {Oh}}, \bibinfo {author} {\bibfnamefont {C.~M.}\ \bibnamefont {Ko}}, \bibinfo
  {author} {\bibfnamefont {S.~H.}\ \bibnamefont {Lee}}, \ and\ \bibinfo
  {author} {\bibfnamefont {S.}~\bibnamefont {Yasui}},\ }\href {\doibase
  10.1103/PhysRevC.79.044905} {\bibfield  {journal} {\bibinfo  {journal} {Phys.
  Rev. C}\ }\textbf {\bibinfo {volume} {79}},\ \bibinfo {pages} {044905}
  (\bibinfo {year} {2009})},\ \Eprint {http://arxiv.org/abs/0901.1382}
  {arXiv:0901.1382 [nucl-th]} \BibitemShut {NoStop}%
\bibitem [{\citenamefont {Plumari}\ \emph {et~al.}(2018)\citenamefont
  {Plumari}, \citenamefont {Minissale}, \citenamefont {Das}, \citenamefont
  {Coci},\ and\ \citenamefont {Greco}}]{Plumari:2017ntm}%
  \BibitemOpen
  \bibfield  {author} {\bibinfo {author} {\bibfnamefont {S.}~\bibnamefont
  {Plumari}}, \bibinfo {author} {\bibfnamefont {V.}~\bibnamefont {Minissale}},
  \bibinfo {author} {\bibfnamefont {S.~K.}\ \bibnamefont {Das}}, \bibinfo
  {author} {\bibfnamefont {G.}~\bibnamefont {Coci}}, \ and\ \bibinfo {author}
  {\bibfnamefont {V.}~\bibnamefont {Greco}},\ }\href {\doibase
  10.1140/epjc/s10052-018-5828-7} {\bibfield  {journal} {\bibinfo  {journal}
  {Eur. Phys. J. C}\ }\textbf {\bibinfo {volume} {78}},\ \bibinfo {pages} {348}
  (\bibinfo {year} {2018})},\ \Eprint {http://arxiv.org/abs/1712.00730}
  {arXiv:1712.00730 [hep-ph]} \BibitemShut {NoStop}%
\bibitem [{\citenamefont {Singha}(2019)}]{Singha:2018cdj}%
  \BibitemOpen
  \bibfield  {author} {\bibinfo {author} {\bibfnamefont {S.}~\bibnamefont
  {Singha}} (\bibinfo {collaboration} {STAR}),\ }\bibfield  {booktitle} {\emph
  {\bibinfo {booktitle} {{Proceedings, 27th International Conference on
  Ultrarelativistic Nucleus-Nucleus Collisions (Quark Matter 2018): Venice,
  Italy, May 14-19, 2018}}},\ }\href {\doibase 10.1016/j.nuclphysa.2018.09.010}
  {\bibfield  {journal} {\bibinfo  {journal} {Nucl. Phys.}\ }\textbf {\bibinfo
  {volume} {A982}},\ \bibinfo {pages} {671} (\bibinfo {year} {2019})},\ \Eprint
  {http://arxiv.org/abs/1807.04771} {arXiv:1807.04771 [nucl-ex]} \BibitemShut
  {NoStop}%
\bibitem [{\citenamefont {Song}\ \emph {et~al.}(2017)\citenamefont {Song},
  \citenamefont {Gou}, \citenamefont {Shao},\ and\ \citenamefont
  {Liang}}]{Song:2017gcz}%
  \BibitemOpen
  \bibfield  {author} {\bibinfo {author} {\bibfnamefont {J.}~\bibnamefont
  {Song}}, \bibinfo {author} {\bibfnamefont {X.-r.}\ \bibnamefont {Gou}},
  \bibinfo {author} {\bibfnamefont {F.-l.}\ \bibnamefont {Shao}}, \ and\
  \bibinfo {author} {\bibfnamefont {Z.-T.}\ \bibnamefont {Liang}},\ }\href
  {\doibase 10.1016/j.physletb.2017.10.012} {\bibfield  {journal} {\bibinfo
  {journal} {Phys. Lett.}\ }\textbf {\bibinfo {volume} {B774}},\ \bibinfo
  {pages} {516} (\bibinfo {year} {2017})},\ \Eprint
  {http://arxiv.org/abs/1707.03949} {arXiv:1707.03949 [hep-ph]} \BibitemShut
  {NoStop}%
\bibitem [{\citenamefont {Zhang}\ \emph {et~al.}(2020)\citenamefont {Zhang},
  \citenamefont {Li}, \citenamefont {Shao},\ and\ \citenamefont
  {Song}}]{Zhang:2018vyr}%
  \BibitemOpen
  \bibfield  {author} {\bibinfo {author} {\bibfnamefont {J.-w.}\ \bibnamefont
  {Zhang}}, \bibinfo {author} {\bibfnamefont {H.-h.}\ \bibnamefont {Li}},
  \bibinfo {author} {\bibfnamefont {F.-l.}\ \bibnamefont {Shao}}, \ and\
  \bibinfo {author} {\bibfnamefont {J.}~\bibnamefont {Song}},\ }\href {\doibase
  10.1088/1674-1137/44/1/014101} {\bibfield  {journal} {\bibinfo  {journal}
  {Chin. Phys.}\ }\textbf {\bibinfo {volume} {C44}},\ \bibinfo {pages} {014101}
  (\bibinfo {year} {2020})},\ \Eprint {http://arxiv.org/abs/1811.00975}
  {arXiv:1811.00975 [hep-ph]} \BibitemShut {NoStop}%
\bibitem [{\citenamefont {Song}\ \emph
  {et~al.}(2020{\natexlab{a}})\citenamefont {Song}, \citenamefont {Shao},\ and\
  \citenamefont {Liang}}]{Song:2019sez}%
  \BibitemOpen
  \bibfield  {author} {\bibinfo {author} {\bibfnamefont {J.}~\bibnamefont
  {Song}}, \bibinfo {author} {\bibfnamefont {F.-l.}\ \bibnamefont {Shao}}, \
  and\ \bibinfo {author} {\bibfnamefont {Z.-t.}\ \bibnamefont {Liang}},\ }\href
  {\doibase 10.1103/PhysRevC.102.014911} {\bibfield  {journal} {\bibinfo
  {journal} {Phys. Rev. C}\ }\textbf {\bibinfo {volume} {102}},\ \bibinfo
  {pages} {014911} (\bibinfo {year} {2020}{\natexlab{a}})},\ \Eprint
  {http://arxiv.org/abs/1911.01152} {arXiv:1911.01152 [nucl-th]} \BibitemShut
  {NoStop}%
\bibitem [{\citenamefont {Gou}\ \emph {et~al.}(2017)\citenamefont {Gou},
  \citenamefont {Shao}, \citenamefont {Wang}, \citenamefont {Li},\ and\
  \citenamefont {Song}}]{Gou:2017foe}%
  \BibitemOpen
  \bibfield  {author} {\bibinfo {author} {\bibfnamefont {X.-r.}\ \bibnamefont
  {Gou}}, \bibinfo {author} {\bibfnamefont {F.-l.}\ \bibnamefont {Shao}},
  \bibinfo {author} {\bibfnamefont {R.-q.}\ \bibnamefont {Wang}}, \bibinfo
  {author} {\bibfnamefont {H.-h.}\ \bibnamefont {Li}}, \ and\ \bibinfo {author}
  {\bibfnamefont {J.}~\bibnamefont {Song}},\ }\href {\doibase
  10.1103/PhysRevD.96.094010} {\bibfield  {journal} {\bibinfo  {journal} {Phys.
  Rev.}\ }\textbf {\bibinfo {volume} {D96}},\ \bibinfo {pages} {094010}
  (\bibinfo {year} {2017})},\ \Eprint {http://arxiv.org/abs/1707.06906}
  {arXiv:1707.06906 [hep-ph]} \BibitemShut {NoStop}%
\bibitem [{\citenamefont {Li}\ \emph {et~al.}(2018)\citenamefont {Li},
  \citenamefont {Shao}, \citenamefont {Song},\ and\ \citenamefont
  {Wang}}]{Li:2017zuj}%
  \BibitemOpen
  \bibfield  {author} {\bibinfo {author} {\bibfnamefont {H.-H.}\ \bibnamefont
  {Li}}, \bibinfo {author} {\bibfnamefont {F.-L.}\ \bibnamefont {Shao}},
  \bibinfo {author} {\bibfnamefont {J.}~\bibnamefont {Song}}, \ and\ \bibinfo
  {author} {\bibfnamefont {R.-Q.}\ \bibnamefont {Wang}},\ }\href {\doibase
  10.1103/PhysRevC.97.064915} {\bibfield  {journal} {\bibinfo  {journal} {Phys.
  Rev.}\ }\textbf {\bibinfo {volume} {C97}},\ \bibinfo {pages} {064915}
  (\bibinfo {year} {2018})},\ \Eprint {http://arxiv.org/abs/1712.08921}
  {arXiv:1712.08921 [hep-ph]} \BibitemShut {NoStop}%
\bibitem [{\citenamefont {Song}\ \emph {et~al.}(2018)\citenamefont {Song},
  \citenamefont {Li},\ and\ \citenamefont {Shao}}]{Song:2018tpv}%
  \BibitemOpen
  \bibfield  {author} {\bibinfo {author} {\bibfnamefont {J.}~\bibnamefont
  {Song}}, \bibinfo {author} {\bibfnamefont {H.-h.}\ \bibnamefont {Li}}, \ and\
  \bibinfo {author} {\bibfnamefont {F.-l.}\ \bibnamefont {Shao}},\ }\href
  {\doibase 10.1140/epjc/s10052-018-5817-x} {\bibfield  {journal} {\bibinfo
  {journal} {Eur. Phys. J.}\ }\textbf {\bibinfo {volume} {C78}},\ \bibinfo
  {pages} {344} (\bibinfo {year} {2018})},\ \Eprint
  {http://arxiv.org/abs/1801.09402} {arXiv:1801.09402 [hep-ph]} \BibitemShut
  {NoStop}%
\bibitem [{\citenamefont {Song}\ \emph
  {et~al.}(2020{\natexlab{b}})\citenamefont {Song}, \citenamefont {Wang},
  \citenamefont {Li}, \citenamefont {Wang},\ and\ \citenamefont
  {Shao}}]{song2020strange}%
  \BibitemOpen
  \bibfield  {author} {\bibinfo {author} {\bibfnamefont {J.}~\bibnamefont
  {Song}}, \bibinfo {author} {\bibfnamefont {X.-f.}\ \bibnamefont {Wang}},
  \bibinfo {author} {\bibfnamefont {H.-h.~L.}\ \bibnamefont {Li}}, \bibinfo
  {author} {\bibfnamefont {R.-q.~W.}\ \bibnamefont {Wang}}, \ and\ \bibinfo
  {author} {\bibfnamefont {F.-l.}\ \bibnamefont {Shao}},\ }\href@noop {} {\
  (\bibinfo {year} {2020}{\natexlab{b}})},\ \Eprint
  {http://arxiv.org/abs/2007.14588} {arXiv:2007.14588 [nucl-th]} \BibitemShut
  {NoStop}%
\bibitem [{\citenamefont {Adamczyk}\ \emph
  {et~al.}(2016{\natexlab{b}})\citenamefont {Adamczyk} \emph
  {et~al.}}]{Adamczyk:2015fum}%
  \BibitemOpen
  \bibfield  {author} {\bibinfo {author} {\bibfnamefont {L.}~\bibnamefont
  {Adamczyk}} \emph {et~al.} (\bibinfo {collaboration} {STAR}),\ }\href
  {\doibase 10.1103/PhysRevC.93.014907} {\bibfield  {journal} {\bibinfo
  {journal} {Phys. Rev.}\ }\textbf {\bibinfo {volume} {C93}},\ \bibinfo {pages}
  {014907} (\bibinfo {year} {2016}{\natexlab{b}})},\ \Eprint
  {http://arxiv.org/abs/1509.08397} {arXiv:1509.08397 [nucl-ex]} \BibitemShut
  {NoStop}%
\bibitem [{\citenamefont {Zhu}(2019)}]{Zhu:2018zqo}%
  \BibitemOpen
  \bibfield  {author} {\bibinfo {author} {\bibfnamefont {Y.}~\bibnamefont
  {Zhu}} (\bibinfo {collaboration} {ALICE}),\ }\bibfield  {booktitle} {\emph
  {\bibinfo {booktitle} {{Proceedings, 39th International Conference on High
  Energy Physics (ICHEP2018): Seoul, Korea, July 4-11, 2018}}},\ }\href
  {\doibase 10.22323/1.340.0441} {\bibfield  {journal} {\bibinfo  {journal}
  {PoS}\ }\textbf {\bibinfo {volume} {ICHEP2018}},\ \bibinfo {pages} {441}
  (\bibinfo {year} {2019})}\BibitemShut {NoStop}%
\bibitem [{\citenamefont {Abelev}\ \emph {et~al.}(2008)\citenamefont {Abelev}
  \emph {et~al.}}]{Abelev:2008jga}%
  \BibitemOpen
  \bibfield  {author} {\bibinfo {author} {\bibfnamefont {B.~I.}\ \bibnamefont
  {Abelev}} \emph {et~al.} (\bibinfo {collaboration} {STAR}),\ }\href {\doibase
  10.1103/PhysRevLett.101.252301} {\bibfield  {journal} {\bibinfo  {journal}
  {Phys. Rev. Lett.}\ }\textbf {\bibinfo {volume} {101}},\ \bibinfo {pages}
  {252301} (\bibinfo {year} {2008})},\ \Eprint {http://arxiv.org/abs/0807.1518}
  {arXiv:0807.1518 [nucl-ex]} \BibitemShut {NoStop}%
\bibitem [{\citenamefont {Abelev}\ \emph
  {et~al.}(2013{\natexlab{a}})\citenamefont {Abelev} \emph
  {et~al.}}]{Abelev:2013cva}%
  \BibitemOpen
  \bibfield  {author} {\bibinfo {author} {\bibfnamefont {B.}~\bibnamefont
  {Abelev}} \emph {et~al.} (\bibinfo {collaboration} {ALICE}),\ }\href
  {\doibase 10.1103/PhysRevLett.111.232302} {\bibfield  {journal} {\bibinfo
  {journal} {Phys. Rev. Lett.}\ }\textbf {\bibinfo {volume} {111}},\ \bibinfo
  {pages} {232302} (\bibinfo {year} {2013}{\natexlab{a}})},\ \Eprint
  {http://arxiv.org/abs/1306.4145} {arXiv:1306.4145 [nucl-ex]} \BibitemShut
  {NoStop}%
\bibitem [{\citenamefont {Baltz}\ and\ \citenamefont
  {Dover}(1996)}]{Baltz:1995tv}%
  \BibitemOpen
  \bibfield  {author} {\bibinfo {author} {\bibfnamefont {A.~J.}\ \bibnamefont
  {Baltz}}\ and\ \bibinfo {author} {\bibfnamefont {C.}~\bibnamefont {Dover}},\
  }\href {\doibase 10.1103/PhysRevC.53.362} {\bibfield  {journal} {\bibinfo
  {journal} {Phys. Rev.}\ }\textbf {\bibinfo {volume} {C53}},\ \bibinfo {pages}
  {362} (\bibinfo {year} {1996})}\BibitemShut {NoStop}%
\bibitem [{\citenamefont {Alt}\ \emph {et~al.}(2008)\citenamefont {Alt} \emph
  {et~al.}}]{Alt:2008iv}%
  \BibitemOpen
  \bibfield  {author} {\bibinfo {author} {\bibfnamefont {C.}~\bibnamefont
  {Alt}} \emph {et~al.} (\bibinfo {collaboration} {NA49}),\ }\href {\doibase
  10.1103/PhysRevC.78.044907} {\bibfield  {journal} {\bibinfo  {journal} {Phys.
  Rev.}\ }\textbf {\bibinfo {volume} {C78}},\ \bibinfo {pages} {044907}
  (\bibinfo {year} {2008})},\ \Eprint {http://arxiv.org/abs/0806.1937}
  {arXiv:0806.1937 [nucl-ex]} \BibitemShut {NoStop}%
\bibitem [{\citenamefont {Sun}\ \emph {et~al.}(2012)\citenamefont {Sun},
  \citenamefont {Wang}, \citenamefont {Song},\ and\ \citenamefont
  {Shao}}]{Sun:2011kj}%
  \BibitemOpen
  \bibfield  {author} {\bibinfo {author} {\bibfnamefont {L.-X.}\ \bibnamefont
  {Sun}}, \bibinfo {author} {\bibfnamefont {R.-Q.}\ \bibnamefont {Wang}},
  \bibinfo {author} {\bibfnamefont {J.}~\bibnamefont {Song}}, \ and\ \bibinfo
  {author} {\bibfnamefont {F.-L.}\ \bibnamefont {Shao}},\ }\href {\doibase
  10.1088/1674-1137/36/1/009} {\bibfield  {journal} {\bibinfo  {journal} {Chin.
  Phys.}\ }\textbf {\bibinfo {volume} {C36}},\ \bibinfo {pages} {55} (\bibinfo
  {year} {2012})},\ \Eprint {http://arxiv.org/abs/1105.0577} {arXiv:1105.0577
  [hep-ph]} \BibitemShut {NoStop}%
\bibitem [{\citenamefont {Lin}\ and\ \citenamefont
  {Molnar}(2003)}]{Lin:2003jy}%
  \BibitemOpen
  \bibfield  {author} {\bibinfo {author} {\bibfnamefont {Z.-w.}\ \bibnamefont
  {Lin}}\ and\ \bibinfo {author} {\bibfnamefont {D.}~\bibnamefont {Molnar}},\
  }\href {\doibase 10.1103/PhysRevC.68.044901} {\bibfield  {journal} {\bibinfo
  {journal} {Phys. Rev.}\ }\textbf {\bibinfo {volume} {C68}},\ \bibinfo {pages}
  {044901} (\bibinfo {year} {2003})},\ \Eprint
  {http://arxiv.org/abs/nucl-th/0304045} {arXiv:nucl-th/0304045 [nucl-th]}
  \BibitemShut {NoStop}%
\bibitem [{\citenamefont {Wang}\ \emph {et~al.}(2020)\citenamefont {Wang},
  \citenamefont {Song}, \citenamefont {Shao},\ and\ \citenamefont
  {Liang}}]{Wang:2019fcg}%
  \BibitemOpen
  \bibfield  {author} {\bibinfo {author} {\bibfnamefont {R.-Q.}\ \bibnamefont
  {Wang}}, \bibinfo {author} {\bibfnamefont {J.}~\bibnamefont {Song}}, \bibinfo
  {author} {\bibfnamefont {F.-L.}\ \bibnamefont {Shao}}, \ and\ \bibinfo
  {author} {\bibfnamefont {Z.-T.}\ \bibnamefont {Liang}},\ }\href {\doibase
  10.1103/PhysRevC.101.054903} {\bibfield  {journal} {\bibinfo  {journal}
  {Phys. Rev. C}\ }\textbf {\bibinfo {volume} {101}},\ \bibinfo {pages}
  {054903} (\bibinfo {year} {2020})},\ \Eprint
  {http://arxiv.org/abs/1911.00823} {arXiv:1911.00823 [hep-ph]} \BibitemShut
  {NoStop}%
\bibitem [{\citenamefont {Cao}\ \emph {et~al.}(2015)\citenamefont {Cao},
  \citenamefont {Qin},\ and\ \citenamefont {Bass}}]{Cao:2015hia}%
  \BibitemOpen
  \bibfield  {author} {\bibinfo {author} {\bibfnamefont {S.}~\bibnamefont
  {Cao}}, \bibinfo {author} {\bibfnamefont {G.-Y.}\ \bibnamefont {Qin}}, \ and\
  \bibinfo {author} {\bibfnamefont {S.~A.}\ \bibnamefont {Bass}},\ }\href
  {\doibase 10.1103/PhysRevC.92.024907} {\bibfield  {journal} {\bibinfo
  {journal} {Phys. Rev. C}\ }\textbf {\bibinfo {volume} {92}},\ \bibinfo
  {pages} {024907} (\bibinfo {year} {2015})},\ \Eprint
  {http://arxiv.org/abs/1505.01413} {arXiv:1505.01413 [nucl-th]} \BibitemShut
  {NoStop}%
\bibitem [{\citenamefont {Sirunyan}\ \emph {et~al.}(2018)\citenamefont
  {Sirunyan} \emph {et~al.}}]{Sirunyan:2017plt}%
  \BibitemOpen
  \bibfield  {author} {\bibinfo {author} {\bibfnamefont {A.~M.}\ \bibnamefont
  {Sirunyan}} \emph {et~al.} (\bibinfo {collaboration} {CMS}),\ }\href
  {\doibase 10.1103/PhysRevLett.120.202301} {\bibfield  {journal} {\bibinfo
  {journal} {Phys. Rev. Lett.}\ }\textbf {\bibinfo {volume} {120}},\ \bibinfo
  {pages} {202301} (\bibinfo {year} {2018})},\ \Eprint
  {http://arxiv.org/abs/1708.03497} {arXiv:1708.03497 [nucl-ex]} \BibitemShut
  {NoStop}%
\bibitem [{\citenamefont {Collaboration}(2019)}]{CMS:2019rjv}%
  \BibitemOpen
  \bibfield  {author} {\bibinfo {author} {\bibfnamefont {C.}~\bibnamefont
  {Collaboration}} (\bibinfo {collaboration} {CMS}),\ }\href@noop {} {\emph
  {\bibinfo {title} {{Search for strong electromagnetic fields in PbPb
  collisions at 5.02 TeV via azimuthal anisotropy of $\mathrm{D^0}$ and
  $\mathrm{\overline{D}^0}$ mesons}}}},\ \bibinfo {type} {Tech. Rep.}\ \bibinfo
  {number} {CMS-PAS-HIN-19-008}\ (\bibinfo {year} {2019})\BibitemShut {NoStop}%
\bibitem [{\citenamefont {Vermunt}(2019)}]{Vermunt:2019ecg}%
  \BibitemOpen
  \bibfield  {author} {\bibinfo {author} {\bibfnamefont {L.}~\bibnamefont
  {Vermunt}} (\bibinfo {collaboration} {ALICE}),\ }\href@noop {} {\enquote
  {\bibinfo {title} {{Measurement of $\Lambda_{\rm c}^{+}$ baryons and ${\rm
  D}_{\rm s}^{+}$ mesons in Pb-Pb collisions with ALICE}},}\ } (\bibinfo {year}
  {2019}),\ \Eprint {http://arxiv.org/abs/1910.11738} {arXiv:1910.11738
  [nucl-ex]} \BibitemShut {NoStop}%
\bibitem [{\citenamefont {Kolb}\ \emph {et~al.}(2000)\citenamefont {Kolb},
  \citenamefont {Sollfrank},\ and\ \citenamefont {Heinz}}]{Kolb:2000sd}%
  \BibitemOpen
  \bibfield  {author} {\bibinfo {author} {\bibfnamefont {P.~F.}\ \bibnamefont
  {Kolb}}, \bibinfo {author} {\bibfnamefont {J.}~\bibnamefont {Sollfrank}}, \
  and\ \bibinfo {author} {\bibfnamefont {U.~W.}\ \bibnamefont {Heinz}},\ }\href
  {\doibase 10.1103/PhysRevC.62.054909} {\bibfield  {journal} {\bibinfo
  {journal} {Phys. Rev.}\ }\textbf {\bibinfo {volume} {C62}},\ \bibinfo {pages}
  {054909} (\bibinfo {year} {2000})},\ \Eprint
  {http://arxiv.org/abs/hep-ph/0006129} {arXiv:hep-ph/0006129 [hep-ph]}
  \BibitemShut {NoStop}%
\bibitem [{\citenamefont {Rapp}\ and\ \citenamefont {van
  Hees}(2010)}]{Rapp:2009my}%
  \BibitemOpen
  \bibfield  {author} {\bibinfo {author} {\bibfnamefont {R.}~\bibnamefont
  {Rapp}}\ and\ \bibinfo {author} {\bibfnamefont {H.}~\bibnamefont {van
  Hees}},\ }in\ \href {\doibase 10.1142/9789814293297_0003} {\emph {\bibinfo
  {booktitle} {{Quark-gluon plasma 4}}}}\ (\bibinfo {year} {2010})\ pp.\
  \bibinfo {pages} {111--206},\ \Eprint {http://arxiv.org/abs/0903.1096}
  {arXiv:0903.1096 [hep-ph]} \BibitemShut {NoStop}%
\bibitem [{\citenamefont {Svetitsky}(1988)}]{Svetitsky:1987gq}%
  \BibitemOpen
  \bibfield  {author} {\bibinfo {author} {\bibfnamefont {B.}~\bibnamefont
  {Svetitsky}},\ }\href {\doibase 10.1103/PhysRevD.37.2484} {\bibfield
  {journal} {\bibinfo  {journal} {Phys. Rev.}\ }\textbf {\bibinfo {volume}
  {D37}},\ \bibinfo {pages} {2484} (\bibinfo {year} {1988})}\BibitemShut
  {NoStop}%
\bibitem [{\citenamefont {Moore}\ and\ \citenamefont
  {Teaney}(2005)}]{Moore:2004tg}%
  \BibitemOpen
  \bibfield  {author} {\bibinfo {author} {\bibfnamefont {G.~D.}\ \bibnamefont
  {Moore}}\ and\ \bibinfo {author} {\bibfnamefont {D.}~\bibnamefont {Teaney}},\
  }\href {\doibase 10.1103/PhysRevC.71.064904} {\bibfield  {journal} {\bibinfo
  {journal} {Phys. Rev.}\ }\textbf {\bibinfo {volume} {C71}},\ \bibinfo {pages}
  {064904} (\bibinfo {year} {2005})},\ \Eprint
  {http://arxiv.org/abs/hep-ph/0412346} {arXiv:hep-ph/0412346 [hep-ph]}
  \BibitemShut {NoStop}%
\bibitem [{\citenamefont {van Hees}\ and\ \citenamefont
  {Rapp}(2005)}]{vanHees:2004gq}%
  \BibitemOpen
  \bibfield  {author} {\bibinfo {author} {\bibfnamefont {H.}~\bibnamefont {van
  Hees}}\ and\ \bibinfo {author} {\bibfnamefont {R.}~\bibnamefont {Rapp}},\
  }\href {\doibase 10.1103/PhysRevC.71.034907} {\bibfield  {journal} {\bibinfo
  {journal} {Phys. Rev.}\ }\textbf {\bibinfo {volume} {C71}},\ \bibinfo {pages}
  {034907} (\bibinfo {year} {2005})},\ \Eprint
  {http://arxiv.org/abs/nucl-th/0412015} {arXiv:nucl-th/0412015 [nucl-th]}
  \BibitemShut {NoStop}%
\bibitem [{\citenamefont {He}\ \emph {et~al.}(2012)\citenamefont {He},
  \citenamefont {Fries},\ and\ \citenamefont {Rapp}}]{He:2011qa}%
  \BibitemOpen
  \bibfield  {author} {\bibinfo {author} {\bibfnamefont {M.}~\bibnamefont
  {He}}, \bibinfo {author} {\bibfnamefont {R.~J.}\ \bibnamefont {Fries}}, \
  and\ \bibinfo {author} {\bibfnamefont {R.}~\bibnamefont {Rapp}},\ }\href
  {\doibase 10.1103/PhysRevC.86.014903} {\bibfield  {journal} {\bibinfo
  {journal} {Phys. Rev.}\ }\textbf {\bibinfo {volume} {C86}},\ \bibinfo {pages}
  {014903} (\bibinfo {year} {2012})},\ \Eprint {http://arxiv.org/abs/1106.6006}
  {arXiv:1106.6006 [nucl-th]} \BibitemShut {NoStop}%
\bibitem [{\citenamefont {Cao}\ \emph {et~al.}(2013)\citenamefont {Cao},
  \citenamefont {Qin},\ and\ \citenamefont {Bass}}]{Cao:2013ita}%
  \BibitemOpen
  \bibfield  {author} {\bibinfo {author} {\bibfnamefont {S.}~\bibnamefont
  {Cao}}, \bibinfo {author} {\bibfnamefont {G.-Y.}\ \bibnamefont {Qin}}, \ and\
  \bibinfo {author} {\bibfnamefont {S.~A.}\ \bibnamefont {Bass}},\ }\href
  {\doibase 10.1103/PhysRevC.88.044907} {\bibfield  {journal} {\bibinfo
  {journal} {Phys. Rev.}\ }\textbf {\bibinfo {volume} {C88}},\ \bibinfo {pages}
  {044907} (\bibinfo {year} {2013})},\ \Eprint {http://arxiv.org/abs/1308.0617}
  {arXiv:1308.0617 [nucl-th]} \BibitemShut {NoStop}%
\bibitem [{\citenamefont {Das}\ \emph {et~al.}(2015)\citenamefont {Das},
  \citenamefont {Scardina}, \citenamefont {Plumari},\ and\ \citenamefont
  {Greco}}]{Das:2015ana}%
  \BibitemOpen
  \bibfield  {author} {\bibinfo {author} {\bibfnamefont {S.~K.}\ \bibnamefont
  {Das}}, \bibinfo {author} {\bibfnamefont {F.}~\bibnamefont {Scardina}},
  \bibinfo {author} {\bibfnamefont {S.}~\bibnamefont {Plumari}}, \ and\
  \bibinfo {author} {\bibfnamefont {V.}~\bibnamefont {Greco}},\ }\href
  {\doibase 10.1016/j.physletb.2015.06.003} {\bibfield  {journal} {\bibinfo
  {journal} {Phys. Lett.}\ }\textbf {\bibinfo {volume} {B747}},\ \bibinfo
  {pages} {260} (\bibinfo {year} {2015})},\ \Eprint
  {http://arxiv.org/abs/1502.03757} {arXiv:1502.03757 [nucl-th]} \BibitemShut
  {NoStop}%
\bibitem [{\citenamefont {Acharya}\ \emph
  {et~al.}(2020{\natexlab{a}})\citenamefont {Acharya} \emph
  {et~al.}}]{Acharya:2019qge}%
  \BibitemOpen
  \bibfield  {author} {\bibinfo {author} {\bibfnamefont {S.}~\bibnamefont
  {Acharya}} \emph {et~al.} (\bibinfo {collaboration} {ALICE}),\ }\href
  {\doibase 10.1016/j.physletb.2020.135225} {\bibfield  {journal} {\bibinfo
  {journal} {Phys. Lett. B}\ }\textbf {\bibinfo {volume} {802}},\ \bibinfo
  {pages} {135225} (\bibinfo {year} {2020}{\natexlab{a}})},\ \Eprint
  {http://arxiv.org/abs/1910.14419} {arXiv:1910.14419 [nucl-ex]} \BibitemShut
  {NoStop}%
\bibitem [{\citenamefont {Adam}\ \emph {et~al.}(2017)\citenamefont {Adam} \emph
  {et~al.}}]{Adam:2017zbf}%
  \BibitemOpen
  \bibfield  {author} {\bibinfo {author} {\bibfnamefont {J.}~\bibnamefont
  {Adam}} \emph {et~al.} (\bibinfo {collaboration} {ALICE}),\ }\href {\doibase
  10.1103/PhysRevC.95.064606} {\bibfield  {journal} {\bibinfo  {journal} {Phys.
  Rev.}\ }\textbf {\bibinfo {volume} {C95}},\ \bibinfo {pages} {064606}
  (\bibinfo {year} {2017})},\ \Eprint {http://arxiv.org/abs/1702.00555}
  {arXiv:1702.00555 [nucl-ex]} \BibitemShut {NoStop}%
\bibitem [{\citenamefont {Adamczyk}\ \emph
  {et~al.}(2016{\natexlab{c}})\citenamefont {Adamczyk} \emph
  {et~al.}}]{Adamczyk:2015lvo}%
  \BibitemOpen
  \bibfield  {author} {\bibinfo {author} {\bibfnamefont {L.}~\bibnamefont
  {Adamczyk}} \emph {et~al.} (\bibinfo {collaboration} {STAR}),\ }\href
  {\doibase 10.1103/PhysRevC.93.021903} {\bibfield  {journal} {\bibinfo
  {journal} {Phys. Rev.}\ }\textbf {\bibinfo {volume} {C93}},\ \bibinfo {pages}
  {021903} (\bibinfo {year} {2016}{\natexlab{c}})},\ \Eprint
  {http://arxiv.org/abs/1506.07605} {arXiv:1506.07605 [nucl-ex]} \BibitemShut
  {NoStop}%
\bibitem [{\citenamefont {Abelev}\ \emph {et~al.}(2009)\citenamefont {Abelev}
  \emph {et~al.}}]{Abelev:2008aa}%
  \BibitemOpen
  \bibfield  {author} {\bibinfo {author} {\bibfnamefont {B.~I.}\ \bibnamefont
  {Abelev}} \emph {et~al.} (\bibinfo {collaboration} {STAR}),\ }\href {\doibase
  10.1103/PhysRevC.79.064903} {\bibfield  {journal} {\bibinfo  {journal} {Phys.
  Rev.}\ }\textbf {\bibinfo {volume} {C79}},\ \bibinfo {pages} {064903}
  (\bibinfo {year} {2009})},\ \Eprint {http://arxiv.org/abs/0809.4737}
  {arXiv:0809.4737 [nucl-ex]} \BibitemShut {NoStop}%
\bibitem [{\citenamefont {Kalinak}(2017)}]{Kalinak:2017xll}%
  \BibitemOpen
  \bibfield  {author} {\bibinfo {author} {\bibfnamefont {P.}~\bibnamefont
  {Kalinak}} (\bibinfo {collaboration} {ALICE}),\ }\bibfield  {booktitle}
  {\emph {\bibinfo {booktitle} {{Proceedings, 2017 European Physical Society
  Conference on High Energy Physics (EPS-HEP 2017): Venice, Italy, July 5-12,
  2017}}},\ }\href {\doibase 10.22323/1.314.0168} {\bibfield  {journal}
  {\bibinfo  {journal} {PoS}\ }\textbf {\bibinfo {volume} {EPS-HEP2017}},\
  \bibinfo {pages} {168} (\bibinfo {year} {2017})}\BibitemShut {NoStop}%
\bibitem [{\citenamefont {Acharya}\ \emph
  {et~al.}(2020{\natexlab{b}})\citenamefont {Acharya} \emph
  {et~al.}}]{Acharya:2019yoi}%
  \BibitemOpen
  \bibfield  {author} {\bibinfo {author} {\bibfnamefont {S.}~\bibnamefont
  {Acharya}} \emph {et~al.} (\bibinfo {collaboration} {ALICE}),\ }\href
  {\doibase 10.1103/PhysRevC.101.044907} {\bibfield  {journal} {\bibinfo
  {journal} {Phys. Rev. C}\ }\textbf {\bibinfo {volume} {101}},\ \bibinfo
  {pages} {044907} (\bibinfo {year} {2020}{\natexlab{b}})},\ \Eprint
  {http://arxiv.org/abs/1910.07678} {arXiv:1910.07678 [nucl-ex]} \BibitemShut
  {NoStop}%
\bibitem [{\citenamefont {Abelev}\ \emph
  {et~al.}(2013{\natexlab{b}})\citenamefont {Abelev} \emph
  {et~al.}}]{Abelev:2013vea}%
  \BibitemOpen
  \bibfield  {author} {\bibinfo {author} {\bibfnamefont {B.}~\bibnamefont
  {Abelev}} \emph {et~al.} (\bibinfo {collaboration} {ALICE}),\ }\href
  {\doibase 10.1103/PhysRevC.88.044910} {\bibfield  {journal} {\bibinfo
  {journal} {Phys. Rev.}\ }\textbf {\bibinfo {volume} {C88}},\ \bibinfo {pages}
  {044910} (\bibinfo {year} {2013}{\natexlab{b}})},\ \Eprint
  {http://arxiv.org/abs/1303.0737} {arXiv:1303.0737 [hep-ex]} \BibitemShut
  {NoStop}%
\bibitem [{\citenamefont {Adams}\ \emph {et~al.}(2004)\citenamefont {Adams}
  \emph {et~al.}}]{Adams:2003xp}%
  \BibitemOpen
  \bibfield  {author} {\bibinfo {author} {\bibfnamefont {J.}~\bibnamefont
  {Adams}} \emph {et~al.} (\bibinfo {collaboration} {STAR}),\ }\href {\doibase
  10.1103/PhysRevLett.92.112301} {\bibfield  {journal} {\bibinfo  {journal}
  {Phys. Rev. Lett.}\ }\textbf {\bibinfo {volume} {92}},\ \bibinfo {pages}
  {112301} (\bibinfo {year} {2004})},\ \Eprint
  {http://arxiv.org/abs/nucl-ex/0310004} {arXiv:nucl-ex/0310004 [nucl-ex]}
  \BibitemShut {NoStop}%
\bibitem [{\citenamefont {Adams}\ \emph {et~al.}(2006)\citenamefont {Adams}
  \emph {et~al.}}]{Adams:2006wk}%
  \BibitemOpen
  \bibfield  {author} {\bibinfo {author} {\bibfnamefont {J.}~\bibnamefont
  {Adams}} \emph {et~al.} (\bibinfo {collaboration} {STAR, STAR RICH}),\
  }\href@noop {} {\  (\bibinfo {year} {2006})},\ \Eprint
  {http://arxiv.org/abs/nucl-ex/0601042} {arXiv:nucl-ex/0601042 [nucl-ex]}
  \BibitemShut {NoStop}%
\bibitem [{\citenamefont {Adamczyk}\ \emph {et~al.}(2017)\citenamefont
  {Adamczyk} \emph {et~al.}}]{Adamczyk:2017iwn}%
  \BibitemOpen
  \bibfield  {author} {\bibinfo {author} {\bibfnamefont {L.}~\bibnamefont
  {Adamczyk}} \emph {et~al.} (\bibinfo {collaboration} {STAR}),\ }\href
  {\doibase 10.1103/PhysRevC.96.044904} {\bibfield  {journal} {\bibinfo
  {journal} {Phys. Rev.}\ }\textbf {\bibinfo {volume} {C96}},\ \bibinfo {pages}
  {044904} (\bibinfo {year} {2017})},\ \Eprint
  {http://arxiv.org/abs/1701.07065} {arXiv:1701.07065 [nucl-ex]} \BibitemShut
  {NoStop}%
\bibitem [{\citenamefont {Adam}\ \emph {et~al.}(2020)\citenamefont {Adam} \emph
  {et~al.}}]{Adam:2019koz}%
  \BibitemOpen
  \bibfield  {author} {\bibinfo {author} {\bibfnamefont {J.}~\bibnamefont
  {Adam}} \emph {et~al.} (\bibinfo {collaboration} {STAR}),\ }\href {\doibase
  10.1103/PhysRevC.102.034909} {\bibfield  {journal} {\bibinfo  {journal}
  {Phys. Rev. C}\ }\textbf {\bibinfo {volume} {102}},\ \bibinfo {pages}
  {034909} (\bibinfo {year} {2020})},\ \Eprint
  {http://arxiv.org/abs/1906.03732} {arXiv:1906.03732 [nucl-ex]} \BibitemShut
  {NoStop}%
\bibitem [{\citenamefont {Abelev}\ \emph
  {et~al.}(2013{\natexlab{c}})\citenamefont {Abelev} \emph
  {et~al.}}]{Abelev:2013xaa}%
  \BibitemOpen
  \bibfield  {author} {\bibinfo {author} {\bibfnamefont {B.~B.}\ \bibnamefont
  {Abelev}} \emph {et~al.} (\bibinfo {collaboration} {ALICE}),\ }\href
  {\doibase 10.1103/PhysRevLett.111.222301} {\bibfield  {journal} {\bibinfo
  {journal} {Phys. Rev. Lett.}\ }\textbf {\bibinfo {volume} {111}},\ \bibinfo
  {pages} {222301} (\bibinfo {year} {2013}{\natexlab{c}})},\ \Eprint
  {http://arxiv.org/abs/1307.5530} {arXiv:1307.5530 [nucl-ex]} \BibitemShut
  {NoStop}%
\bibitem [{\citenamefont {Abelev}\ \emph {et~al.}(2014)\citenamefont {Abelev}
  \emph {et~al.}}]{ABELEV:2013zaa}%
  \BibitemOpen
  \bibfield  {author} {\bibinfo {author} {\bibfnamefont {B.~B.}\ \bibnamefont
  {Abelev}} \emph {et~al.} (\bibinfo {collaboration} {ALICE}),\ }\href
  {\doibase 10.1016/j.physletb.2014.05.052, 10.1016/j.physletb.2013.11.048}
  {\bibfield  {journal} {\bibinfo  {journal} {Phys. Lett.}\ }\textbf {\bibinfo
  {volume} {B728}},\ \bibinfo {pages} {216} (\bibinfo {year} {2014})},\
  \bibinfo {note} {[Erratum: Phys. Lett.B734,409(2014)]},\ \Eprint
  {http://arxiv.org/abs/1307.5543} {arXiv:1307.5543 [nucl-ex]} \BibitemShut
  {NoStop}%
\end{thebibliography}%

\appendix
\section{$v_{2}$ of pion and kaon \label{app_v2_piK}}

\begin{figure}[h]
\centering{}\includegraphics[scale=0.4]{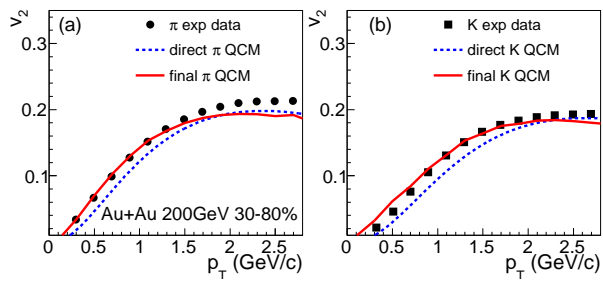}\caption{$v_{2}$ of pion and kaon in Au+Au collisions at $\sqrt{s_{NN}}=200$
GeV for 30-80\% centrality. Symbols are experimental data \cite{Adamczyk:2015ukd}.
Dashed lines are results of $v_{2}$ directly applying EVC formulas.
Solid lines are results of $u+\bar{d}/\bar{s}\to\pi/K+X$ process
after considering the decay contribution of other hadrons. }
\label{fig_b1}
\end{figure}

Because the masses of pion and kaon are small, the production of pion
and kaon is not suitably described by the direct combination of constituent
quarks and antiquarks. To reconcile the mass mismatch, we adopt a
naive treatment \cite{Gou:2017foe} which provides the good description
for the $p_{T}$ distribution of pions and that of kaons. We consider
the processes such as $u+\bar{d}\to\pi+X$ for pion production and
$u+\bar{s}\to K+X$ for kaon production. Here $X$ is some soft degrees
of freedom. For simplicity, we identify $X$ as soft pions. As an
example, using the extracted $v_{2}$ of $u$ and $s$ quarks in Au+Au
collisions at $\sqrt{s_{NN}}=200$ GeV for 30-80\% centrality, we
calculate $v_{2}$ of the directly-produced pions and kaons by above
processes and then consider the decay contribution of other hadrons.
We show results as solid lines, marked as ``final $\pi/K$ QCM'',
in Fig. \ref{fig_b1} and compare with experimental data \cite{Adamczyk:2015ukd}.
We see that $v_{2}$ of pion and kaon in the low $p_{T}$ range ($p_{T}\lesssim2$
GeV/c) can be described by $v_{2}$ of quarks that is extracted from
baryons and $\phi$.

As a contrast, we also calculate $v_{2}$ of pions and that of kaons
by direct EVC formulas, i.e., $v_{2,\pi}\left(p_{T}\right)=2v_{2,u}\left(p_{T}/2\right)$
and $v_{2,K}\left(p_{T}\right)=v_{2,u}\left(\frac{1}{1+r}p_{T}\right)+v_{2,s}\left(\frac{r}{1+r}p_{T}\right)$.
We present results as dashed lines in Fig. \ref{fig_b1}, marked as
``direct $\pi/K$ QCM''. Comparing to solid lines, we see the important
effect of extra $X$ in pion and kaon production and that of resonance
decays.

\section{the influence of two-kaon coalescence on $p_{T}$ spectrum of $\phi$\label{sec:appA}}

Starting from Eqs. (\ref{eq:fptphi_mi}) and (\ref{eq:fptphi_bi}),
the $p_{T}$ distributions of $\Omega$ and $\phi$ under equal-velocity
combination have
\begin{align}
f_{\Omega}\left(p_{T}\right) & =\kappa_{\Omega}f_{s}^{3}\left(p_{T}/3\right),\label{eq:fphi}\\
f_{\phi}\left(p_{T}\right) & =\kappa_{\phi}f_{s}^{2}\left(p_{T}/2\right)\label{eq:fOmega}
\end{align}
from which we get a quark number scaling property
\begin{equation}
f_{\Omega}^{1/3}\left(3p_{T}\right)=\kappa_{\phi,\Omega}f_{\phi}^{1/2}\left(2p_{T}\right),\label{eq:qns_fpt}
\end{equation}
where $\kappa_{\phi,\Omega}=\kappa_{\Omega}^{1/3}/\kappa_{\phi}^{1/2}$
is independent of $p_{T}$. We take $f_{s}\left(p_{T}\right)=f_{\bar{s}}\left(p_{T}\right)$
at LHC energies.

As the discussion of $\phi$ elliptic flow at LHC energies in Sec.
\ref{sec:v2_LHC}, the coalescence of two kaons may be another contribution
to $\phi$ production. The $p_{T}$ spectrum of $\phi$ by the coalescence
of two charged kaons with equal velocity is
\begin{equation}
f_{\phi,KK}\left(p_{T}\right)\propto f_{K^{+}}\left(p_{T}/2\right)f_{K^{-}}\left(p_{T}/2\right)\label{eq:phi_kpkm}
\end{equation}
or by that of two neutral kaons is
\begin{equation}
f_{\phi,KK}\left(p_{T}\right)\propto f_{K^{0}}\left(p_{T}/2\right)f_{\bar{K}^{0}}\left(p_{T}/2\right)\propto\left[f_{K_{s}^{0}}\left(p_{T}/2\right)\right]^{2}.\label{eq:phi_2ks0}
\end{equation}
\ins{We use experimental data of $K^{\pm}$ and $K_{s}^{0}$ in central heavy-ion collisions to calculate results of two-kaon coalescence by Eqs.~(\ref{eq:phi_kpkm}) and (\ref{eq:phi_2ks0}) and compare calculation results with experimental data of $\phi$ in the same collision centrality. Fig. \ref{fig_a1} shows results and comparisons at four collision energies. The centrality at each collision energy is selected by the condition that experimental data of $K^{\pm}$, $K_{s}^{0}$ and $\phi$ are all available.  }
We see that the spectra of two-kaon coalescence are almost parallel to those
of $\phi$ for $p_{T,\phi}\lesssim4$ GeV/c. This indicates that two-kaon
coalescence does not change the shape of $\phi$ distribution in this
$p_{T}$ range. Therefore, it also does not break the quark number
scaling property Eq. (\ref{eq:qns_fpt}) in the range $p_{T,\phi}\lesssim$
4 GeV/c, equivalently, in the range $p_{T,s}\lesssim2$ GeV/c dominated
by soft or thermal strange quarks.
\begin{center}
\begin{figure}[h]
\centering{}\includegraphics[scale=0.4]{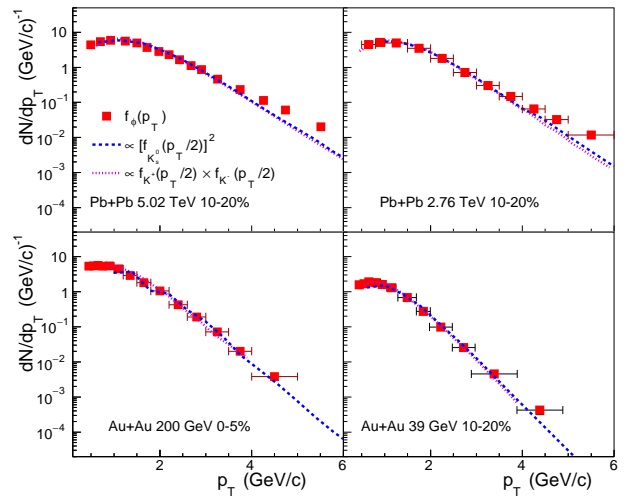}\caption{The $p_{T}$ spectra by the coalescence of two kaons with equal velocity,
which are compared with those of $\phi$. Data of $\phi$ are taken
from Refs.~\cite{Acharya:2019qge,Adam:2017zbf,Adamczyk:2015lvo,Abelev:2008aa}
    and those of kaons \ins{used to calculate two-kaon coalescence} are taken from Refs.~\cite{Kalinak:2017xll,Acharya:2019yoi,Abelev:2013vea,Adams:2003xp,Adams:2006wk,Adamczyk:2017iwn,Adam:2019koz,Abelev:2013xaa}. }
\label{fig_a1}
\end{figure}
\par\end{center}

\begin{figure}[h]
\centering{}\includegraphics[scale=0.3]{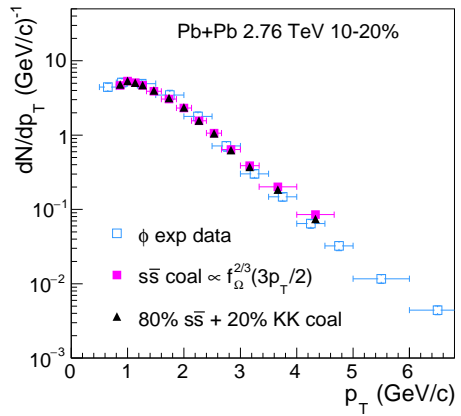}\caption{The $p_{T}$ spectrum of $\phi$ calculated by pure $s\bar{s}$ combination
and that by mixture of $s\bar{s}$ combination and $KK$ coalescence
in Pb+Pb collisions at $\sqrt{s_{NN}}=$ 2.76 TeV for 10-20\% centrality.
Results are compared with data of $\phi$ \cite{Adam:2017zbf}.}
\label{fig_a2}
\end{figure}

In the range $p_{T}\gtrsim4$ GeV/c, see top panels in Fig. \ref{fig_a1},
the spectra of two-kaon coalescence are steeper than those of $\phi$
to a certain extent. Therefore including this contribution will make
$\phi$ spectrum softer than that formed purely by the strange quark
combination. Here, we take a simple case as an illustration, i.e.,
80\% of final-state $\phi$ comes from the direct strange quark combination
and the remaining 20\% comes from two-kaon coalescence. Fig. \ref{fig_a2}
shows our calculations and the comparison with data of $\phi$ in
Pb+Pb collisions at $\sqrt{s_{NN}}=$ 2.76 TeV for 10-20\% centrality
\cite{Adam:2017zbf}. Firstly, we present the results of pure strange
quark combination at hadronization, the solid squares, which can be
calculated using the data of $\Omega$ \cite{ABELEV:2013zaa} by
the scaling property Eq. (\ref{eq:qns_fpt}),
\begin{equation}
f_{\phi,s\bar{s}}\left(p_{T}\right)=\kappa_{\phi,\Omega}^{-2}\,f_{\Omega}^{2/3}\left(3p_{T}/2\right).
\end{equation}
We see that they are in good agreement with data of $\phi$ for $p_{T}\lesssim$
4 GeV/c and the last point at $p_{T}=4.3$ GeV/c is higher than the
$\phi$ datum to a certain extent. Then, we consider the contribution
of two-kaon coalescence and results are shown as up-triangles. Data
of kaons are taken from Ref. \cite{Abelev:2013vea}. We see that
the results for $p_{T}\lesssim$ 3.5 GeV/c are almost unchanged compared
with those of pure strange quark combination. The last two points
at $p_{T}=$3.7 and 4.3 GeV/c are decreased to a certain extent and
are closer to the data of $\phi$.

\end{document}